\documentclass{aa}
\usepackage[dvips]{graphicx}
\usepackage{longtable,lscape}

\begin{document}
 
\title{New periodic variable stars coincident with {\em ROSAT} sources 
discovered using SuperWASP}
 
\titlerunning{WASP/{\em ROSAT} variable stars}
 
\author{A.J. Norton\inst{1}, P.J. Wheatley\inst{2}, R.G. West\inst{3},
C.A. Haswell\inst{1}, R.A. Street\inst{4},
A. Collier Cameron\inst{5}, D.J. Christian\inst{4}, W.I. Clarkson\inst{6}, 
B. Enoch\inst{1}, M. Gallaway\inst{1},
C. Hellier\inst{7}, K. Horne\inst{5}, J. Irwin\inst{8}, S.R. Kane\inst{9}, 
T.A. Lister\inst{7}, J.P. Nicholas\inst{2}, N. Parley\inst{10}, 
D. Pollacco\inst{4}, R. Ryans\inst{4}, I. Skillen\inst{11}, 
D.M. Wilson\inst{7}}
 
\authorrunning{Norton et al}
 
\offprints{A.J. Norton, a.j.norton@open.ac.uk}
 
\institute{Department of Physics and Astronomy, The Open University,
           Walton Hall, Milton Keynes MK7 6AA, U.K.
\and
           Department of Physics, University of Warwick, Coventry CV4 7AL, U.K.
\and      
           Department of Physics and Astronomy, University of Leicester,
           Leicester LE1 7RH, U.K.
\and
           Astrophysics Research Centre, Main Physics Building, School of
           Mathematics \& Physics, Queen's University, University Road, 
           Belfast BT7 1NN, U.K.
\and
           School of Physics and Astronomy, University of St. Andrews,
           North Haugh, St. Andrews, Fife KY16 9SS, U.K.
\and
           Space Telescope Science Institute, 3700 San Martin Drive, 
           Baltimore, MD 21218, U.S.A.
\and 
           Astrophysics Group, School of Chemistry \& Physics, Keele
           University, Staffordshire ST5 5BG, U.K.
\and
           Institute of Astronomy, University of Cambridge, Madingly Road,
           Cambridge CB3 0HA, U.K.
\and
           Department of Physics, University of Florida, Gainesville,
           FL 32611-8440, U.S.A.
\and
           Planetary \& Space Sciences Research Institute, The Open University,
           Walton Hall, Milton Keynes MK7 6AA, U.K.
\and
           Isaac Newton Group of Telescopes, Apartado de Correos 321, 
           E-38700 Santa Cruz de la Palma, Tenerife, Spain}

\date{Accepted 21 Feb 2007;
      Received 12 Jan 2007}
 
\abstract{We present optical lightcurves of 428 periodic 
variable stars coincident with {\em ROSAT} X-ray sources, detected 
using the first run of the SuperWASP photometric survey. Only 68 of these 
were previously recognised as periodic variables. A further 30 of these 
objects are previously known pre-main sequence stars, for which we detect a 
modulation 
period for the first time. Amongst the newly identified periodic variables, 
many appear to be close eclipsing binaries, their X-ray emission is presumably 
the result of RS CVn type behaviour. Others are probably BY Dra stars, 
pre-main sequence stars and other rapid rotators displaying enhanced coronal 
activity. A number of previously catalogued pulsating variables (RR Lyr stars 
and Cepheids) coincident with X-ray sources are also seen, but we show that
these are likely to be misclassifications. We identify four objects which are 
probable low mass eclipsing binary stars, based on their very red colour 
and light curve morphology.

\keywords{stars: binaries: eclipsing -- stars: rotation -- 
stars: variables: general}}

\maketitle

\section{Introduction}

The SuperWASP project (Pollacco et al. 2006) is a wide field photometric 
survey designed to search for transiting exoplanets  and other signatures of 
variability on timescales from minutes to months. In its first run during 
2004, SuperWASP-N on La Palma was operated between May and September with five 
cameras, each of which has a $7.8^{\circ}\times 7.8^{\circ}$ field of view. 
The fields surveyed in 2004 comprise a strip of sky centred at declination 
$~+28^{\circ}$ and extending to all right ascensions (excluding the galactic 
plane). Coverage is not uniform though, with some regions of sky better 
sampled than others. The resulting area covered was $\sim 10,000$ square 
degrees and comprised over 300,000 raw images. Photometry for all objects 
detected was extracted using a 2.5 pixel aperture ($34^{\prime \prime}$ 
radius). As a result, the project produced unfiltered (white light) 
lightcurves of over 6.7 million stars in the magnitude range V$\sim 8-15$, 
totalling 12.9 billion data points.  Because 
of the wide-field nature of the images, systematic errors in the lightcurves 
are present, but are largely removed using the SysREM algorithm from Tamuz 
et al. (2005). Plots showing the RMS precision of our data as a function of 
SuperWASP V magnitude, both before and after application of the SysREM 
algorithm, are shown in Collier Cameron et al. (2006a).

The 2004 survey allowed the identification of $\sim 100$ transiting exoplanet 
candidates, which we have reported in a series of papers (Christian et al. 
2006, Clarkson et al. 2007, Kane et al. in preparation, Lister et al. 2007, 
Street et al. 2007, West et al. in preparation). The first confirmed planets resulting 
from this are reported by Collier Cameron et al. (2006b). These data also uncovered huge 
numbers of variable stars, many of which are previously 
unidentified. In this paper we discuss a small subset of these, namely those 
that are coincident with {\em ROSAT} X-ray sources. Our reasons for 
concentrating on the subset of variable sources which are X-ray sources are 
twofold. First, variable objects coincident with 
X-ray sources are likely to yield significant numbers of astrophysically
important and interesting objects. Secondly, this provides a manageable 
number of objects with which to demonstrate the efficiency of SuperWASP for 
detecting variable objects other than transiting exoplanets. SuperWASP 
provides a significant increase in time coverage as compared to many single 
object studies and smaller photometric campaigns.

\section{Period searching}

The positions of objects in the SuperWASP archive are derived from the USNO-B1
catalogue (Monet et al. 2003). These positions were cross correlated 
against the {\em ROSAT} sky survey (1RXS; Voges et al. 1999, 2000) and pointed 
phase (2RXP; ROSAT 2000) catalogues, 
taking the uncertainty in position of each {\em ROSAT} source as 
its $3\sigma$ error radius, or 10$^{\prime \prime}$, whichever was larger.
This resulted in 4,562 matches between the positions of SuperWASP
objects from our 2004 Northern hemisphere run and the positions of 
{\em ROSAT} sources. From this set of cross identifications, 
3,558 SuperWASP lightcurves have more than 100 data points, and so were 
deemed to be suitable for period searching.  A purpose written period search 
code was then run on these to identify periodic variables, 
coincident with X-ray sources.

The period search comprised two techniques. A {\sc clean}ed power spectrum 
was calculated, using the variable gain implementation of H. Lehto, and the 
strongest peaks within it identified. A period folding analysis was
also performed, searching over periods from 20 minutes to half the lightcurve
length in each case. Binning the data into 20 phase bins at each trial period, 
we calculated the reduced chi-squared of the folded lightcurve with respect
to its mean flux and also the sum of the reduced chi-squareds of the data
within each phase bin with respect to the mean flux in that bin. The most 
likely periods were then taken as those which maximised the difference 
between these two chi-squared values. This jointly minimised the 
dispersion within each phase bin and maximised the dispersion between the 
phase bins, in order to identify likely periods. Only those periods found 
in common between the {\sc clean}ed power spectrum and period folding 
technique (within a tolerance of 1\%) were 
recorded. It was noted that a few systematic effects remain in the SuperWASP 
lightcurves, sometimes resulting in spurious periods being identified due 
to night-to-night variations. As a result, we ignored periods within 1\% 
of one day and fractions thereof (i.e. 1d/2, 1d/3, 1d/4, etc). We also 
rejected around 20 objects where several sources within a few arcminutes of 
each other each displayed similar lightcurves with similar long term (tens 
of days) periodicities. These are believed to be artefacts due to remaining 
systematic errors in the extracted data. After this there remained 516 
SuperWASP sources in this set for which periodic variability was identified, 
with periods ranging from less than 3h to more than 50d. 

88 of these variable sources were noted to be duplicates of other stars in 
the list, recorded with slightly different positions (typically within 
10 arcseconds) due to the large pixel size of the SuperWASP cameras. These
arise where there are multiple USNO-B1 objects within a few arcseconds of 
each other, and the SuperWASP data resulting from different images are 
variously assigned to one of this small set of objects. 
Encouragingly, we detected the {\em same} period in the multiple  
SuperWASP lightcurves in each case. Most of these duplicates consisted of 
just two SuperWASP lightcurves corresponding to the same object, but in a 
couple of cases there were as many as five SuperWASP lightcurves corresponding
to the same object, each with slightly different positions but showing the 
same periodicity. After removing these duplicates, there remained 428 unique 
objects showing periodic variability in their SuperWASP lightcurves and 
coincident with {\em ROSAT} X-ray sources. 

\section{Results}

The positions of the 428 periodic SuperWASP objects were cross-correlated
against the SIMBAD database. In order to account for the $34^{\prime \prime}$ 
radius photometry aperture used for our data, we searched for all 
objects in SIMBAD within twice this distance of the nominal SuperWASP position,
and identified the most likely source of the variable signal in each case.
As a result we identified 68 sources that have been previously recorded 
as periodic variable stars, of which 66 have periods given in the literature. 
These objects comprise 47 listed in the General Catalogue of Variable Stars, 
17 discovered by the ROTSE (Robotic Optical Transient Search Experiment) 
survey (Akerlof et al. 2000), 2 discovered by the SAVS (Semi-Automatic 
Variability Search) survey (Maciejewski et al. 2004) and 2 objects (HD170451 
and SAO46441) recently identified as W UMa type eclipsing binaries but not 
yet assigned GCVS designations. This set of 68 objects also includes two
identified eclipsing binaries (KW Com and V1011 Her) whose periods appear
not to have been previously published.

The details of these 68 objects are listed in Table 1; their folded 
SuperWASP lightcurves and power spectra are shown in Figure 1. Phase zero 
for each of the folded lightcurves is set at 2004 January 1st 00:00UT 
(i.e. HJD 2453005.5). The columns of Table 1
are as follows: 1. The number of SuperWASP objects which are duplicates, 
recorded with slightly different positions but displaying the same period, 
where this number is greater than one. 2. The SuperWASP identifier in the form 
`1SWASP Jhhmmss.ss+ddmmss.s'; the position encoded in this identifier will
be identical to the position of the corresponding object in the 
USNO B1 catalogue. 3. The {\em ROSAT} identifier either from
the 1RXS or 2RXP catalogues.  4. The period in days as derived from the 
SuperWASP lightcurve. 5. The mean SuperWASP magnitude, defined as 
$-2.5\log_{10}(F/10^6)$ where $F$ is the mean SuperWASP flux in microVegas; 
it is a pseudo-V magnitude which is comparable to the Tycho V magnitude. 
6. and 7. The B1 and R1 
magnitudes respectively from the USNO catalogue (Monet et al. 2003). 8. A 
previously recorded name of the object. 9. The astronomical classification
of the object. 10. The previously recorded period of the object. 11. A 
reference to the previous period determination. 

The 68 previously classified periodic variables consist of 13 pre-main
sequence stars, 
10 Algol type (EA) eclipsing binaries (4 of which are also RS CVn stars), 
5 $\beta$ Lyrae type (EB) eclipsing binaries, 10 W UMa type (EW) eclipsing 
binaries, 6 BY Dra systems (5 of which also show UV Cet  type behaviour, one 
of which has a white dwarf companion), 5 RS CVn systems (4 of which are also 
Algol type eclipsing binaries), 2 RR Lyrae stars, 15 Cepheid variables
(13 of which were classified by the ROTSE project and one by the SAVS survey), 
one semi-regular pulsator,  3 cataclysmic variable stars, 
one supersoft source and one low mass X-ray binary.

The periods we have determined from our SuperWASP data for these
known objects are generally in good agreement with previously published
values (see Table 1). Where the periods differ significantly, we are confident
that our determination is the more reliable measurement, owing to the
better sampling of our data. In a couple of cases,
for instance, we detect clear periodicities which are close to {\em half} 
that of objects discovered by ROTSE and claimed to be Cepheid variable stars.
In another case, the cataclysmic variable PX And, the period we detect is the
disc precession period rather than the binary orbital period of 0.14635d.

The remaining 360 objects comprise newly identified periodic 
variable stars which are also X-ray sources. Their details are listed in 
Table 2 and their SuperWASP lightcurves and power spectra are also shown in 
Figure 1. The columns of Table 2 are essentially the same as those in Table 1, 
without the previously determined periods and references, but with
the addition of the spectral type where this is recorded in {\em SIMBAD}.
Many of the objects in Table 2 are anonymous stars, with only a Hubble Space 
Telescope Guide Star Catalog or Tycho Catalog designation. Where objects have 
a designation other than one of these catalogue numbers, that is listed
in Table 2. The period distribution of all 428 objects is shown in 
Figure 2.

In two cases (1SWASP J170033.82+200134.1 and 1SWASP J222229.09+281439.1)
the SuperWASP lightcurves are double valued at all phases. In each case the 
objects are double stars (see Table 2) and the anomalous lightcurves
are undoubtedly the result of one of the two stars (the non-variable one)
sometimes appearing within the photometry aperture and sometimes not, so 
offsetting the mean brightness for a subset of the datapoints.

We also note that some folded lightcurves (e.g. that of 1SWASP 
J141630.88+265525.1) show regular `chunks' of data (7 in this case) such that 
the measured period is close to the same integer number of days (i.e. the 
period is 6.9998d in the case of this object). However, although the phase
coverage is uneven, these lightcurves 
cover many cycles of variation (121 days duration in the case of this object) 
and the period is reliable.  The pattern seen is a 
result of the modulation period being close to an integer number of days and 
the sampling of the object repeating at the same time of night over many 
weeks. The longest period accepted for any object is less than half the data
length in each case.

\section{Discussion}

\subsection{Positional coincidence}

As noted earlier, since the SuperWASP pixel size is relatively large 
(13.7$^{\prime \prime}$~pixel$^{-1}$), the 2.5 pixel extraction aperture for 
photometry corresponds to 34$^{\prime \prime}$ in radius. Given that the 
{\em ROSAT} sources can have positional uncertainties of up to tens of 
arcseconds, there is clearly the likelihood of chance positional coincidences 
between SuperWASP objects and catalogued X-ray sources. 

The sky area covered by our 2004 Northern hemisphere observations is 
about 10$^4$ square degrees, and we have lightcurves of 6,713,217 objects
from this SuperWASP run, of which 5,271,091 have more than 100 data points.
The mean separation between nearest neighbours is therefore about 
140$^{\prime \prime}$ for the total set of SuperWASP objects. There 
are 14,616 {\em ROSAT} sources that fall within 
the sky area covered by our 2004 Northern hemisphere SuperWASP run, 
including 3,826 from 1RXS and 10,790 from 2RXP. (Many of these will, however, 
be duplicates between the two catalogues.) The mean error radius of the 
{\em ROSAT} positions is 18$^{\prime \prime}$. Hence, there is a 
5.3\% chance of a given {\em ROSAT} source coinciding with one of our 
SuperWASP sources, purely at random. The fact that we find 4,562 matches, 
rather than the $\sim 770$ matches that would result from chance alone, 
suggests that at least $(4562 - 770) / 4562 = 83\%$ of the positional 
coincidences are genuinely the result of X-ray emission from SuperWASP 
objects. At least 300 of the newly identified periodic variable stars 
identified here are therefore likely to be actual X-ray emitters.

\subsection{Previously known periodic variable stars}

The list of known variable stars in the sample considered here contains 
five accreting binary stars -- the X-ray binary Her X-1, the supersoft X-ray 
source QR And, and the three cataclysmic variables PX And, V795 Her and DQ Her.
These latter three are all magnetic CVs to some extent, and so have 
enhanced X-ray emission as a result. 

Apart from this, the X-ray emission seen in the other objects with 
previously  known periods is generally a result of stellar coronal activity
(e.g. Rosner, Golub \& Vaiana 1985; Hartmann \& Noyes 1987). The key feature 
linking their X-ray emission is rapid rotation causing enhanced magnetic 
fields through the dynamo mechanism. This is particularly evident in the
13 pre-main sequence stars which we detect whose periods have been previously 
determined (e.g. from the COYOTE campaigns of Bouvier et al. 1993, 1995, 
1997). These are young, rapidly rotating stars, and all those detected here 
are in the Taurus-Auriga star forming complex. We note that the observed 
photometric modulation periods of pre-main sequence stars are due to the 
presence of star spots, and that these can therefore change with time as spots 
appear and disappear at different latitudes. This will give rise to different 
modulation periods at different epochs if the stars rotate differentially 
(Neuhauser et al. 1995).

Another manifestation of coronal X-ray emission due to rapid rotation is in 
RS CVn stars. These are detached  binary systems in which the rotation of 
the two components is locked to the orbital period of typically just a few 
days (Hall 1976; Strassmeier et al. 1993). One of the stars is usually a 
K sub-giant, and it is this star with 
its deep convection zone which develops a strong magnetic field and 
enhanced stellar coronal activity. Through optical selection effects, many
RS CVn systems are detected as eclipsing binaries, and indeed we see that 
four of the five known RS CVn systems in this sample are of the Algol type,
i.e. detached eclipsing binaries.

BY Dra stars are also well-represented amongst the systems with 
previously known periods. These are dwarf K or M stars showing emission 
lines and believed to be rapid rotators (Alekseev 2000; Strassmeier et al. 
1993). Many of them show flare star behaviour and so are classed as 
UV Cet type stars too (Gershberg et a 1999). The periods
we see in the previously known systems of these types are almost certainly 
the rotation periods of the star. One of the previously catalogued BY Dra 
stars is V1092 Tau, which is a wide binary system containing a rapidly 
rotating K2V star and a hot white dwarf (Jeffries, Burleigh \& Robb 1996).

Finally, we apparently see a surprising number of pulsating variable stars 
as X-ray sources. This includes both RR Lyr type, which are A or F giant stars 
with periods typically less than a day (Smith 1995), and $\delta$~Cep type 
(Cepheids) which are super-giants with periods of around 1 to 100 days 
(e.g. Turner \& Burke 2002). Most of the apparent Cepheids recovered
here were discovered by the ROTSE project. This detected 201 Cepheids in 
2000 square degrees of sky coverage (Akerlof et al. 2000), giving a mean 
separation between them of 3.15$^{\circ}$. There is therefore a chance of 
less than 0.001\% that one of these ROTSE Cepheids would coincide with one 
of the {\em ROSAT} sources falling within the SuperWASP survey area. The fact 
that we find 13 ROTSE Cepheids coincident with {\em ROSAT} sources suggests 
that these objects are indeed X-ray emitters. However, whether they are 
actually Cepheid variables is not so certain.

Akerlof et al. (2000) classified ROTSE objects as Cepheids on the basis of 
having sinusoidal lightcurves and periods in the range 1 to 50 days. However,
there is no real evidence that these are likely to be Cepheids rather than
other variables such as RS CVn systems. Objects such as ROTSE1 
J172339.92+352759.3 and ROTSE1 J184633.30+485435.3
which we recover here are 11th magnitude stars with periods of $\sim 24$~d 
and $\sim 5$~d respectively. If they really were Cepheids, the 
period-luminosity relationship (Feast \& Catchpole 1997) would place them at 
distances of 24.5~kpc and 11~kpc. Cepheids are Population I objects and 
therefore mostly lie in the Galactic disk. Since none of the objects we have 
considered here are in the Galactic plane, these distances are unrealistic. 
It is likely that {\em none} of the supposed ROTSE Cepheids we have detected 
as coincident with {\em ROSAT} sources are in fact correctly classified.
The possibility of X-ray emission from pulsating stars was raised by 
Bejgman \& Stepanov (1981) although there is little observational evidence
in support of this other than a recent detection by {\em Chandra} of 
X-rays from the Cepheid Polaris (Evans et al. 2006). However, the observed 
X-ray to optical luminosity ratio of Polaris is many orders of magnitude 
smaller than the corresponding values of the supposed ROTSE Cepheids selected 
here (Engle, Guinan \& Evans, 2006), providing further evidence of their
misclassification.

In addition to discounting the ROTSE objects as true pulsating variables, 
the classification of SAVS J022708+342319 as a Cepheid is also uncertain
(Maciejewski et al. 2004) and the RR Lyr variable HR Aur may be an active 
binary rather than a pulsating star (Loomis \& Schmidt 1989). This leaves
only V845 Her as a potential Cepheid with X-ray emission, but this 
appears to have H$\alpha$ in emission (Schmidt et al. 2004b). This too may be 
a sign of coronal activity and hence a misclassification.

\subsection{Newly identified periodic variable stars}

Amongst the newly identified periodic variable stars coincident with 
{\em ROSAT} sources, the majority are likely to be X-ray sources as 
a result of their rapid rotation. Indeed, 15 objects are previously
identified as BY Dra stars, UV Cet stars, or other miscellaneous flare
stars. We also see a further 30 previously catalogued pre-main sequence stars 
in this sample, which are likely to be rapid rotators too. We emphasise
though that none of these 30 objects have previously been reported as showing 
coherent periodic variability. Within the set of 43 known young stars 
reported here (i.e. 13 with previously determined periods and 30 measured
here for the first time), the 
distribution of periods found is quite broad: 7 have periods shorter than 1~d, 
10 have periods between 1~d and 2~d, 13 between 2~d and 4~d, and 13 between 
4~d and 10~d. We anticipate that the remaining, unclassified objects will 
contain further examples of these various classes of stars displaying 
rotational modulation. 

Significant numbers of the newly identified objects are clearly eclipsing 
binaries with periods from a few hours to a few days. We note though that in 
many cases, the period we have identified will be half the binary period, 
particularly in the case of W UMa type variables which display two minima of 
comparable depth. The newly identified X-ray emitting eclipsing binaries 
appear to include Algol type (e.g. 1SWASP 175540.63+372516.0 and
1SWASP J180331.30+080836.3), $\beta$~Lyr type (e.g. 1SWASP J005101.78+200824.4
and 1SWASP J160248.22+252038.2) and W UMa type (e.g. 1SWASP J021208.77+270818.2
and 1SWASP J133538.39+491406.1) variables, as well as others which display 
more unusual morphology (e.g. 1SWASP J180207.45+183044.2). Many of these 
eclipsing binaries will contain tidally locked stars, so their components 
will also be rapid rotators displaying RS CVn type behaviour. This is likely 
to be the source of the X-ray emission in these cases too. 

\subsection{Flux ratios and modulation amplitudes}

Given that X-ray activity in the majority of sources we have detected
is expected to be linked to rapid rotation, one might expect to see an 
anti-correlation between X-ray emission
and modulation period in our data. The modulation period may be the 
rotational period of a star, or a binary period, but tidal locking 
will make these periods identical for many close binaries, so preserving the
correlation.

Previous studies have indeed shown strong correlations
between X-ray emission and rotation period (e.g. Walter \& Bowyer 1981;
Pallavicini et al. 1981) but for very rapid rotation the X-ray luminosity is 
found to saturate at around 0.1\% of the bolometric luminosity (e.g. 
Vilhu \& Walter 1987; Wheatley 1998). Pizzolato et al. (2003) show that 
saturation occurs at rotation periods between 2 and 10\,d depending on 
stellar mass.

In Figure 3 we plot ratios of X-ray to optical flux against our
measured modulation periods.  $F_X$ is defined as the {\it ROSAT} 
count rate and $F_O$ as the mean SuperWASP flux in microVegas. 
We do not attempt to calculate the more usual ratio of X-ray to bolometric 
luminosity because in most cases we do not have a sufficiently reliable 
measure of colour to estimate the spectral type and hence bolometric 
luminosity.

Figure 3 shows very little dependence of X-ray emission on modulation
period. This indicates that, if the bulk of our sample are coronal
emitters, they must be in the saturated regime. The lack of an obvious
decrease in X-ray to optical flux even at periods longer than 10\,d
indicates that our sample is probably dominated by low mass stars 
($M<~0.6\,M_\odot$, Pizzolato et al. 2003). We find a very weak 
anti-correlation with $(F_{\rm X}/F_{\rm O})\propto(P/{\rm day})^{-0.14}$ 
and a value for the Pearson correlation coefficient squared of
$R^2=0.02$. This may be due to the contribution of binaries containing
giants and subgiants (Dempsey et al. 1993).

We further note that the known rapidly rotating isolated stars (BY Dra type, 
UV Cet type, pre-main sequence stars etc) tend to lie at higher X-ray to 
optical flux ratios 
in this diagram, whilst the binary stars (eclipsing binaries and RS CVn stars) 
tend to lie at lower flux ratios. The relatively low observed X-ray to optical 
flux ratios of the RS CVn stars, when compared with the rapidly rotating single
stars, may be due to the fact that we have not carried out any bolometric
correction. The RS CVn stars will typically be K sub-giants, whilst the 
isolated rapid rotators are mostly M dwarfs, so their bolometric corrections
will be different.

The five accreting binaries follow a 
different trend of increasing X-ray to optical flux ratio with period. In 
this case their X-ray emission is clearly not a result of enhanced coronal 
activity induced by rapid rotation. The cataclysmic variable PX And is detected
here at its disc precession period (4.437d), if it were instead plotted
at the position of its orbital period (0.146d) it would lie on the same
trend as the other four accreting binaries in Figure 3. 

Figure 4 shows a histogram of the X-ray to optical flux ratios in this
sample of 428 objects, compared with the flux ratios of the other 
SuperWASP objects which are coincident with {\em ROSAT} objects but did
not yield a modulation period from the period searching. Interestingly, the 
non-periodic sample shows a spread to larger X-ray to optical flux ratios 
than the periodic sample. These appear to be mostly coincident with 
galaxies or AGN, which are not expected to be periodic variables. However
this sample will also contain periodic sources that are not strongly
modulated, and others where the modulation period is not well sampled 
by the SuperWASP observations.

Figure 5 shows the modulation amplitude for our sample of 428 objects
plotted as a function of modulation period. Here we see no correlation, other 
than to note that 
eclipsing binaries and RS CVn stars generally appear with higher modulation
amplitudes than do the isolated rapidly rotating stars. Since most of the 
newly identified objects lie at lower amplitudes, this might indicate that
the majority of these are likewise rapid rotators, rather than eclipsing
binaries. However, we should also be mindful of selection effects which 
will mean that most of the previously identified eclipsing binaries will tend
to be those with the deepest eclipses.

\subsection{Colours}

It is also instructive to plot the colours of this sample of objects, as this
yields several interesting candidates for further investigation. 
Figure 6 shows the SuperWASP V -- 2MASS K colour versus the 2MASS J -- H
colour of the objects. The solid line shows the approximate locus of the 
main sequence from around A0 to M6, for zero reddening. Every magnitude of V 
band extinction will shift this line upwards by 0.12 in (J--H) and to the 
right by 0.92 in (V--K), according to the extinction law of Wegner (1994). 
It is likely that the slight offset between the locus of the data points and 
the plotted main sequence is a result of the non-standard V magnitude 
calculated from the SuperWASP unfiltered flux, and does not indicate any
significant trend in these objects. The fact that most lie close to this
zero-reddening main sequence suggests that the majority are relatively 
nearby objects. We note that four of the five accreting binaries (i.e. 
all except DQ Her) are amongst the bluest objects on  this plot, whilst 
many of the pre-main sequence stars are amongst the reddest. 

Looking first at the blue end of the colour-colour diagram, the two anonymous 
SuperWASP objects lying closest to the known accreting 
binaries are 1SWASP J132426.35+303314.2 (HD116635) 
with $P=3.35$~d and 1SWASP J153633.39+271029.2 (SAO83906) with $P=1.34$~d. 
Although neither have high X-ray to optical flux ratios, these objects  
warrant further investigation as potential accreting compact binaries. They
are much bluer than their spectral classifications listed on SIMBAD (F and G 
respectively) would suggest. 

The one previously classified rotational variable lying just above the 
accreting binaries with (V--K)= 0.45 and (J--H) = 0.34 is 
1SWASP J034433.95+395948.0 which is positionally coincident with
an anonymous pre-main sequence star listed on SIMBAD. The SuperWASP lightcurve 
clearly 
shows this object to be an eclipsing binary star with a period of 0.2888~d.
Given its blue colour this is likely to be mis-classified and not a pre-main
sequence star after all.

Moving to the red end of the colour-colour diagram, the one unclassified 
object lying above the main sequence amongst the very red pre-main sequence 
stars is 1SWASP J033025.95+310217.9 with (V--K)$=4.63$, (J--H)$=0.84$ and 
$P=2.2308$~d. As this lies close to the previously catalogued pre-main sequence
stars in the Taurus-Auriga star forming complex, this is likely to be another 
example of this class of objects, but previously uncatalogued.

The reddest of the previously classified binary stars is KW Com, with (V--K) 
= 4.3, which is listed in the GCVS as an eclipsing binary, although no period 
has previously been published. Its SuperWASP lightcurve is indistinguishable 
from many of the rotational lightcurves of the various pre-main sequence stars 
we have detected, and we therefore suggest it has been mis-classified in the 
GCVS and is really a young star displaying rotational modulation.

Although most of the unclassified SuperWASP objects lying at the red end of 
the colour-colour plot appear to be further examples of young stars displaying
rotational modulation, there are some exceptions.  In particular we 
note that there are four unclassified SuperWASP objects with (V--K)$>3.4$,
which therefore lie amongst the M stars, and which have lightcurves with 
a morphology that is suggestive of eclipsing binary stars.  These are 
1SWASP J142004.68+390301.5 with $P=0.3693$~d and (V--K)=4.02, 
1SWASP J022050.85+332047.6 with $P=0.1926$~d and (V--K)=3.93, 
1SWASP J220041.59+271513.5 with $P=0.5235$~d and (V--K)=3.61, and
1SWASP J224355.18+293647.6 with $P=0.4443$~d and (V--K)=3.49. These
are likely to be low mass eclipsing binaries, and representatives 
of a previously very poorly sampled population.

\section{Conclusions}

We have demonstrated the effectiveness of the SuperWASP survey for
detecting photometric modulation on timescales of hours to weeks, in 
objects within the magnitude range $\sim 8 - 15$. As a result we have 
recovered the previously identified periodicities in 68
known variable stars coincident with {\em ROSAT} X-ray sources, and 
identified a modulation period for the first time in 360 more.
By selecting on objects which are coincident with X-ray sources we have
identified eclipsing binary stars and those showing rotational modulation, 
as well as picking out a few known accreting compact binary stars. We have
shown that several previously catalogued pulsating variables coincident with 
{\em ROSAT} sources are likely to be misclassifications. Finally we have 
identified 4 objects as potential low mass eclipsing binaries on the basis 
of their lightcurve morphology and colours.

\begin{acknowledgements}

The WASP project is funded and operated by Queen's University Belfast, the 
Universities of Keele, St. Andrews and Leicester, the Open University, the
Isaac Newton Group, the Instituto de Astrofisica de Canarias, the South
African Astronomical Observatory and by PPARC. 

This research has made extensive use of the SIMBAD database, operated at CDS, 
Strasbourg, France. We thank Harry Lehto for use of his implementation of 
the 1D {\sc clean} algorithm.

\end{acknowledgements}

\onecolumn

\clearpage


 \begin{figure*}[t]
\begin{center}
\includegraphics[scale=0.5,angle=-90]{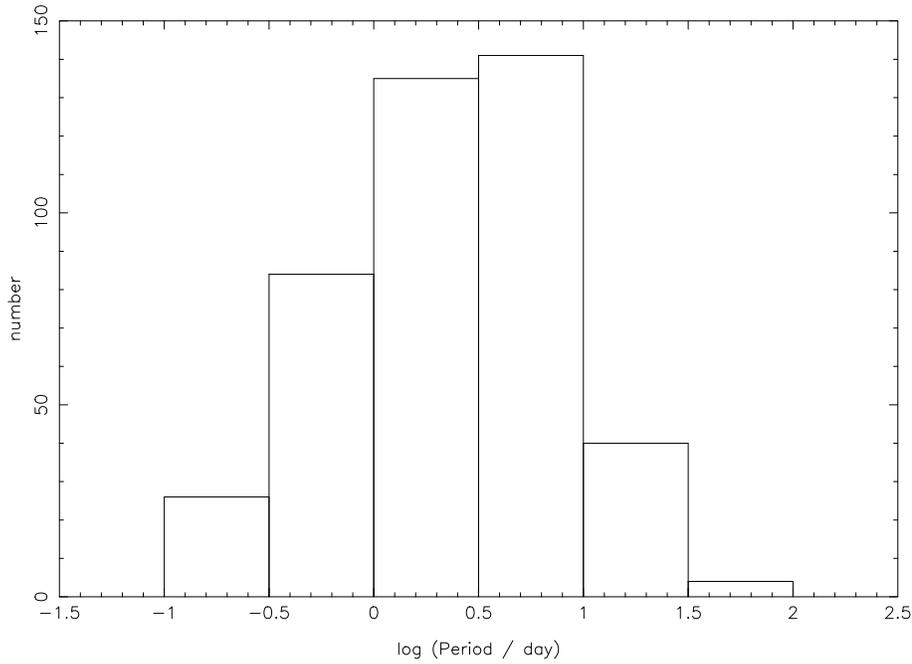}
\caption{The period distribution of our sample of 428 periodic variable
stars coincident with {\em ROSAT} sources.}
\end{center}
 \end{figure*}

 \begin{figure*}[t]
\begin{center}
\includegraphics[scale=0.5,angle=-90]{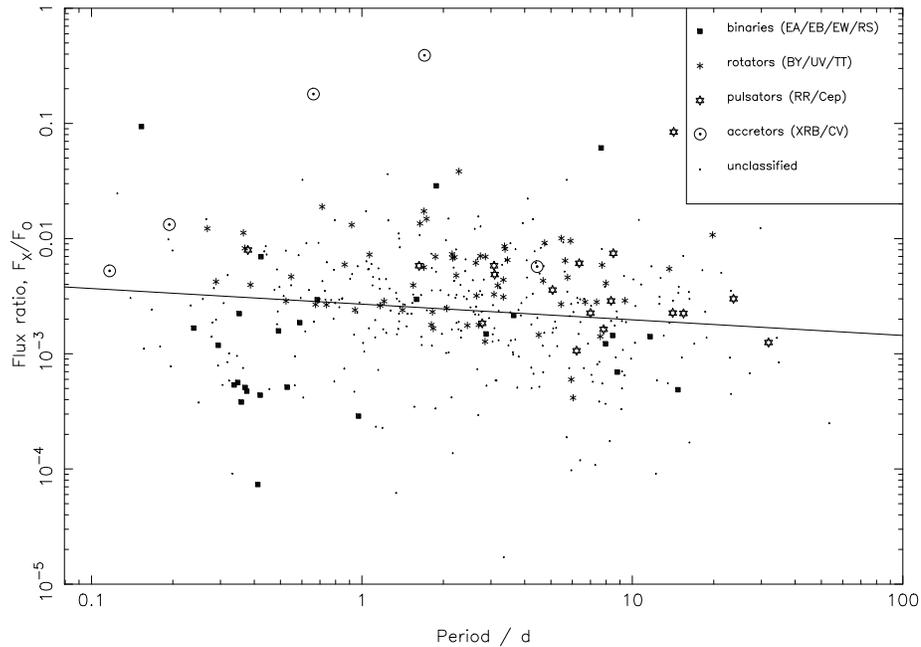}
\caption{The X-ray to optical flux ratio (calculated as the mean {\em ROSAT}
count rate divided by SuperWASP mean flux) plotted against measured
period for our sample of 428 periodic variable stars. The line shows the 
best-fit correlation defined by $F_{\rm X}/F_{\rm O} \propto 
(P/{\rm day})^{-0.14}$.
The symbols representing previously classified objects are shown in the inset
key. `Binaries' include Algol, $\beta$ Lyr and W UMa type eclipsing binaries,
as well as RS CVn stars; `rotators' include BY Dra, UV Cet and pre-main
sequence stars; `pulsators'
include those stars catalogued as either RR Lyr or $\delta$ Cep variables; 
`accretors' include X-ray binaries and cataclysmic variables.}
\end{center}
 \end{figure*}

 \begin{figure*}[t]
\begin{center}
\includegraphics[scale=0.5,angle=-90]{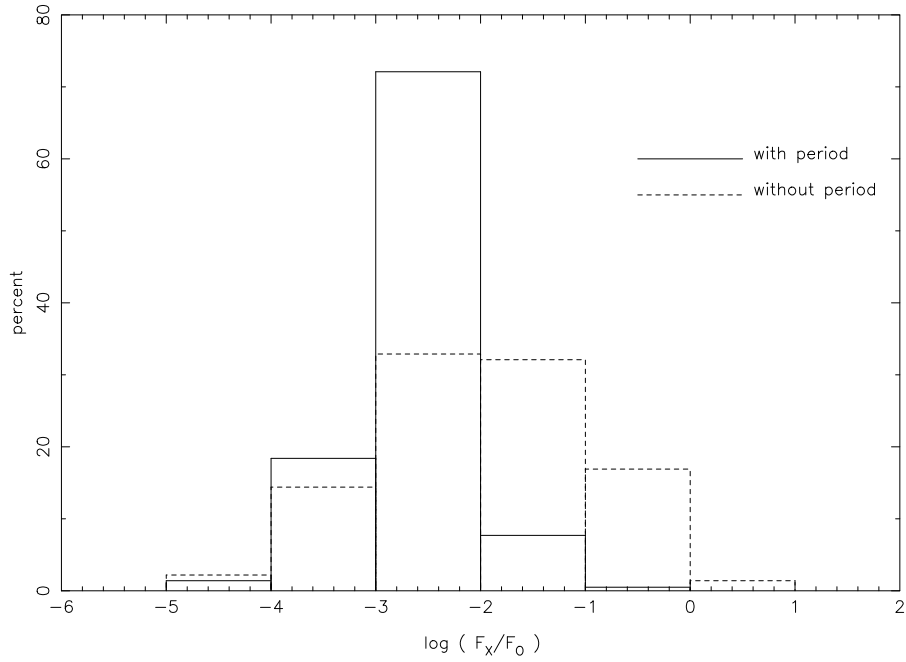}
\caption{The distribution of X-ray to optical flux ratio for all SuperWASP
objects coincident with {\em ROSAT} sources. Both the
sample of 428 objects displaying a period, and the remaining objects for
which no period was found, are shown. Percentages are shown with respect
to the individual sample size in each case.}
\end{center}
 \end{figure*}

 \begin{figure*}[t]
\begin{center}
\includegraphics[scale=0.5,angle=-90]{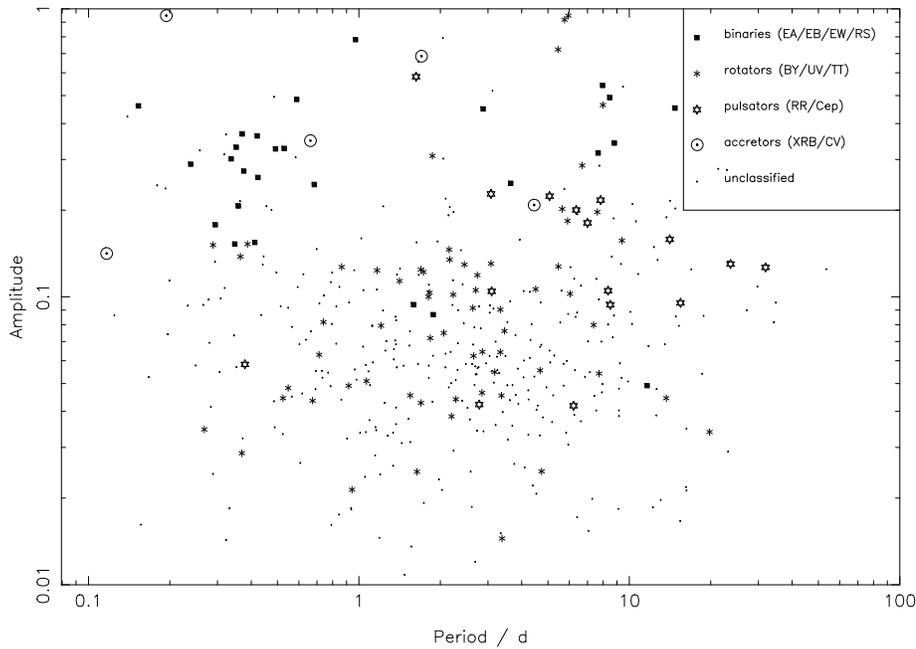}
\caption{The fractional modulation amplitude (calculated as the maximum flux
minus minimum flux, divided by the maximum flux, measured from the 
folded and binned lightcurve) plotted against 
measured period for our sample of 428 periodic variable stars. No correlation 
is apparent. Symbols are as for Figure 3.}
\end{center}
 \end{figure*}

 \begin{figure*}[t]
\begin{center}
\includegraphics[scale=0.5,angle=-90]{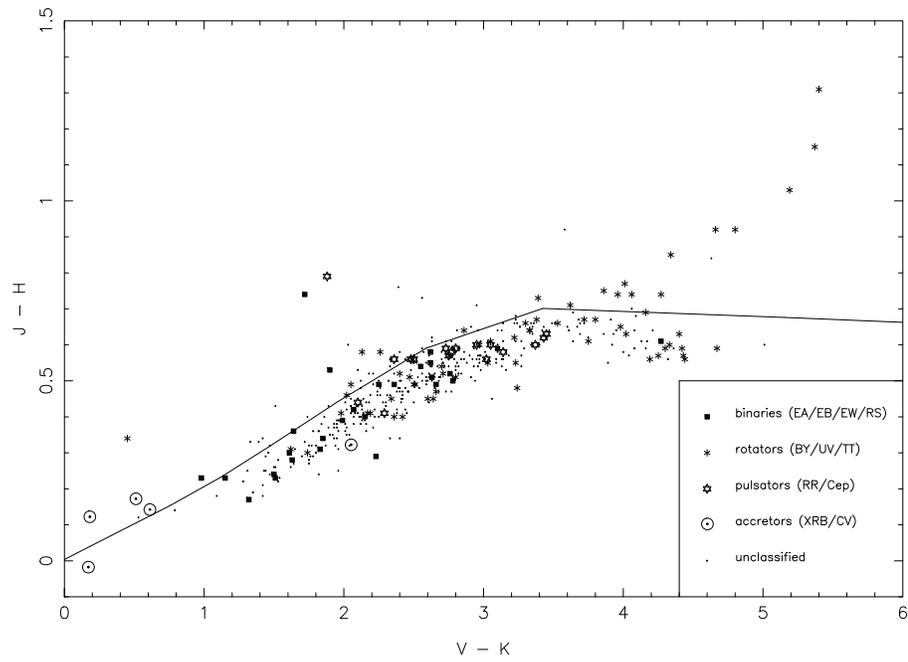}
\caption{The SuperWASP V -- 2MASS K colour plotted as a function of the 
2MASS J -- K colour for our sample of 428 variable stars. The accreting
binaries are the amongst the bluest objects in this diagram, whilst some
of the pre-main sequence stars are amongst the reddest. Symbols are as for 
Figure 3. The
solid line shows the main sequence, for zero reddening, from A0 to late M.}
\end{center}
 \end{figure*}

\tiny
\begin{landscape}
\begin{longtable}{lllrrrrllll}
\caption{SuperWASP objects showing periodic variability, coincident with 
{\em ROSAT} sources and previously classified as periodic variable stars.}\\ \hline \hline
N & SuperWASP ID & {\em ROSAT} ID & Period/d & SW V & B1 & R1 & Name & Type & Period/d & Period ref. \\ \hline
\endfirsthead
\caption{continued.}\\ \hline \hline
N & SuperWASP ID & {\em ROSAT} ID & Period/d & SW V & B1 & R1 & Name & Type & Period/d & Period ref. \\ \hline
\endhead
\hline
\endfoot
  & J001949.91+215652.3 & 2RXP J001949.8+215655 & 0.6604 & 12.73 & 12.38 & 11.86 & QR And & SuperSoft Source & 0.660458 & Chrastina et al. 2006  \\
  & J003005.79+261726.5 & 2RXP J003006.0+261722 & 4.4366 & 14.88 & 15.38 & 14.69 & PX And & SW Sex & 4.8 & Stanishev et al. 2002  \\
2 & J004117.33+342516.8 & 1RXS J004117.7+342513 & 2.8615 & 10.14 & 11.05 & 9.46 & QT And & BY Dra/UV Cet & 1.536 & Robb 1995  \\
3 & J005845.91+320620.3 & 1RXS J005846.2+320627 & 0.6840 & 10.65 & 10.43 & 9.45 & CP Psc & $\beta$ Lyr  & 0.684014 & Selam 2004  \\
  & J012728.88+290618.5 & 1RXS J012729.4+290622 & 0.4916 & 10.91 & 11.57 & 10.58 & SAVS J012728+290618 & W UMa  & 0.491495 & Maciejewski et al. 2004  \\
  & J015757.78+374822.4 & 2RXP J015757.2+374819 & 0.4122 & 11.53 & 11.77 & 11.23 & QX And & W UMa  & 0.4118165 & Pribulla et al. 2003  \\
  & J020033.73+275319.2 & 1RXS J020034.3+275303 & 0.9716 & 9.19 & 8.85 & 8.50 & X Tri & Algol  & 0.9715352 & Samus et al. 2004  \\
  & J022708.42+342320.4 & 1RXS J022709.2+342334 & 3.0990 & 11.77 & 12.82 & 11.97 & SAVS J022708+342319 & Cepheid & 3.1078 & Maciejewski et al. 2004  \\
4 & J035705.82+283751.5 & 1RXS J035706.3+283805 & 0.3647 & 11.87 & 12.81 & 10.92 & V1092 Tau & BY Dra/UV Cet + WD & 0.3646 & Jeffries et al. 1996 \\
2 & J040651.34+254128.4 & 1RXS J040651.0+254159 & 1.6955 & 11.87 & 13.28 & 10.93 & V1195 Tau & PMS star  & 1.73 & Bouvier et al. 1997  \\ 
2 & J040909.74+290130.2 & 1RXS J040909.7+290134 & 2.6538 & 10.79 & 11.56 & 10.24 & V1197 Tau & PMS star  & 2.74 & Bouvier et al. 1997 \\ 
  & J041412.91+281212.3 & 2RXP J041413.2+281212 & 3.0802 & 10.60 & 10.05 & 7.49  & V773 Tau  & PMS star  & 3.43 & Bouvier et al. 1993 \\ 
  & J041447.96+275234.7 & 1RXS J041448.0+275246 & 7.3808 & 11.73 & 14.08 & 10.95 & V1098 Tau & PMS star  & 7.2  & Bouvier et al. 1995  \\ 
  & J041522.91+204417.0 & 2RXP J041523.5+204416 & 1.8232 & 10.84 & 11.23 & 10.12 & V1199 Tau & PMS star  & 1.83 & Bouvier et al. 1997\\ 
  & J041831.10+282716.0 & 2RXP J041831.1+282718 & 1.8704 & 10.98 & 9.96 & 8.45 & V410 Tau & PMS star  & 1.872 & Shevchenko et al. 1999 \\ 
  & J041941.26+274948.3 & 1RXS J041941.7+274953 & 5.6425 & 12.25 & 13.75 & 11.14 & V1070 Tau & PMS star  & 5.64 & Bouvier et al. 1993 \\ 
  & J042448.18+264315.8 & 1RXS J042448.4+264320 & 3.1805 & 11.03 & 14.00 & 10.88 & V1201 Tau & PMS star & 1.89 & Bouvier et al. 1997 \\ 
  & J043116.85+215025.2 & 1RXS J043117.1+215010 & 2.7045 & 10.95 & 11.67 & 10.36 & V1202 Tau & PMS star  & 2.71 & Bouvier et al. 1997  \\ 
  & J043310.02+243343.7 & 2RXP J043310.3+243345 & 2.7429 & 12.08 & 13.91 & 11.30 & V830 Tau & PMS star  & 2.75 & Shevchenko et al. 1999  \\ 
  & J044104.70+245106.1 & 1RXS J044105.2+245107 & 5.4654 & 12.38 & 14.25 & 11.65 & IW Tau & PMS star  & 5.45 & Bouvier et al. 1993  \\ 
  & J045536.96+301755.1 & 1RXS J045536.6+301752 & 2.2321 & 11.24 & 12.34 & 10.55 & V396 Aur & PMS star  & 2.24 & Bouvier et al. 1993  \\ 
2 & J045602.45+302051.6 & 1RXS J045601.7+302058 & 9.3898 & 11.34 & 14.39 & 10.99 & V397 Aur & PMS star  & 10.1 & Bouvier et al. 1995 \\ 
  & J063111.02+305615.9 & 1RXS J063112.5+305614 & 1.6277 & 11.56 & 12.52 & 10.57 & HR Aur & RR Lyr  & 1.627777 & Loomis \& Schmidt 1989  \\
  & J112541.63+423449.6 & 1RXS J112540.3+423449 & 0.4236 & 12.46 & 13.41 & 11.54 & BS UMa & Algol & 0.437016 & Meinunger \& Wenzel 1968  \\
3 & J114749.04+351335.2 & 1RXS J114749.2+351339 & 0.3519 & 10.95 & 11.83 & 10.32 & KM UMa & $\beta$ Lyr  & 0.351862 & Escola-Sirisi \& Garcia-Melendo 1999  \\
  & J123156.02+353015.6 & 1RXS J123156.4+353011 & 0.1530 & 15.25 & 15.54 & 14.36 & DI CVn & $\beta$ Lyr & 0.30599 & Akerlof et al. 2000   \\ 
  & J125336.27+224735.4 & 1RXS J125336.5+224742 & 1.8832 & 13.95 & 16.14 & 13.33 & KW Com & eclipsing binary & ? &  \\
2 & J125740.26+351330.1 & 2RXP J125740.4+351324 & 3.3664 & 10.32 & 12.33 &  9.73 & BF CVn & BY Dra/UV Cet  & 3.17 & Pettersen 1980  \\
  & J130133.01+283754.2 & 1RXS J130133.9+283753 & 3.6411 & 10.10 & 9.51 & 8.15 & UX Com & RS CVn/Algol  & 3.642583 & Hall \& Kreiner 1980  \\
  & J133146.61+291636.6 & 1RXS J133146.9+291631 & 0.2683 & 11.19 & 12.94 & 11.80 & DG CVn & BY Dra/UV Cet  & 0.10835 & Robb et al. 1999  \\
  & J141630.88+265525.1 & 2RXP J141631.0+265527 & 6.9998 & 10.55 & 11.19 & 9.98 & ROTSE1 J141630.86+265524.8 & Cepheid & 7.0209 & Akerlof et al. 2000  \\
  & J142019.60+275856.5 & 1RXS J142019.9+275851 & 7.8284 & 11.13 & 12.34 & 10.50 & ROTSE1 J142019.68+275856.1 & Cepheid & 7.8677 & Akerlof et al. 2000  \\
  & J143230.53+504940.6 & 1RXS J143230.5+504944 & 0.4207 &  9.56 &  9.94 & 9.14	 & EF Boo & W UMa type & 0.420608 & Samec et al. 1999 \\
2 & J143819.50+363225.7 & 1RXS J143820.6+363230 & 0.2389 & 10.56 & 11.78 & 10.50 & GK Boo & Algol & 0.23889 & Akerlof et al. 2000  \\
  & J160650.69+271634.5 & 1RXS J160651.0+271631 & 0.5889 & 10.66 & 11.04 & 10.19 & TW CrB & $\beta$ Lyr  & 0.58879 & Akerlof et al. 2000  \\
2 & J162506.55+300225.8 & 1RXS J162506.5+300218 & 14.1106 & 10.03 & 11.41 & 9.64 & ROTSE1 J162506.53+300225.7 & Cepheid & 28.4301 & Akerlof et al. 2000  \\
2 & J162510.31+405334.2 & 2RXP J162510.0+405340 & 15.4579 & 12.88 & 9.70 & 12.45 & V845 Her & Cepheid & 15.5 & Schmidt et al. 2004a \\ 
  & J162817.29+371124.1 & 2RXP J162817.0+371116 & 0.3370 & 11.70 & 11.89 & 11.14 & ROTSE1 J162817.23+371123.9 & W UMa  & 0.33704 & Akerlof et al. 2000  \\
  & J165547.87+351057.6 & 2RXP J165548.1+351100 & 0.3582 & 10.39 & 10.83 & 9.97 & V829 Her & W UMa & 0.3581 & Robb 1989  \\
2 & J165749.80+352032.8 & 2RXP J165749.4+352033 & 1.7002 & 13.83 & 14.89 & 13.88 & HZ Her/Her X-1 & X-ray binary & 1.700175 & Deeter et al. 1991  \\
  & J170121.84+420949.9 & 1RXS J170121.5+420939 & 0.3702 & 9.66 & 10.12 & 9.18 & SAO 46441 & W UMa & 0.37015 & anon. 2005 (IBVS 5600)  \\
  & J170420.20+392858.7 & 1RXS J170420.7+392909 & 1.5929 & 11.70 & 12.43 & 10.97 & ROTSE1 J170420.17+392858.0 & W UMa  & 1.593435 & Akerlof et al. 2000  \\
  & J170757.96+291915.0 & 1RXS J170757.4+291922 & 3.0816 & 12.04 & 12.75 & 11.01 & ROTSE1 J170757.98+291914.9 & Cepheid & 3.077795 & Akerlof et al. 2000  \\
  & J171256.18+333119.2 & 2RXP J171256.0+333121 & 0.1165 & 13.00 & 12.41 & 12.69 & V795 Her & novalike & 0.11588 & Mironov et al. 1983  \\
  & J171330.96+232026.4 & 1RXS J171331.0+232021 & 2.7852 & 10.64 & 11.45 & 10.22 & ROTSE1 J171330.93+232026.3 & Cepheid & 2.75599 & Akerlof et al. 2000  \\
  & J172339.79+352757.2 & 1RXS J172339.8+352757 & 23.6758 & 11.65 & 12.09 & 10.81 & ROTSE1 J172339.92+352759.3 & Cepheid & 23.675575 & Akerlof et al. 2000  \\
  & J172425.27+493837.2 & 1RXS J172425.4+493831 & 0.5295 & 9.62 & 10.09 & 9.29 & V878 Her & $\beta$ Lyr & 0.529478 & Kaiser et al. 1996  \\
  & J172927.24+352404.9 & 1RXS J172927.2+352402 & 14.2202 & 11.81 & 13.55 & 10.96 & ROTSE1 J172927.28+352403.0 & Cepheid & 13.7084 & Akerlof et al. 2000  \\
  & J174311.06+334948.9 & 1RXS J174311.5+334950 & 6.3627 & 11.74 & 12.82 & 10.57 & ROTSE1 J174311.03+334948.1 & Cepheid & 6.33071 & Akerlof et al. 2000  \\
  & J174605.24+312104.5 & 1RXS J174604.8+312058 & 0.3789 & 12.43 & 13.94 & 11.44 & ROTSE1 J174605.20+312104.1 & RR Lyr  & 0.378799 & Akerlof et al. 2000  \\
  & J175838.51+220846.7 & 1RXS J175838.2+220835 & 7.9603 & 9.90 & 9.24 & 8.01 & MM Her & RS CVn/Algol & 7.960322 & Hall \& Kreiner 1980  \\
  & J180147.31+273910.4 & 1RXS J180147.5+273918 & 6.2201 & 10.42 & 11.06 & 9.74 & ROTSE1 J180147.49+273907.4 & Cepheid & 6.07072 & Akerlof et al. 2000  \\
  & J180213.87+470112.3 & 1RXS J180214.5+470112 & 0.2944 & 10.94 & 11.82 & 10.34 & ROTSE1 J180213.83+470112.3 & W UMa  & 0.29433 & Akerlof et al. 2000  \\
  & J180730.28+455131.9 & 2RXP J180730.0+455136 & 0.1936 & 15.16 & 12.99 & 11.46 & DQ Her & nova & 0.19362 & Zhang et al. 1995  \\
  & J180853.52+370707.3 & 1RXS J180853.5+370702 & 8.3429 & 11.87 & 13.43 & 13.66 & ROTSE1 J180853.47+370708.1 & Cepheid & 8.214765 & Akerlof et al. 2000  \\
  & J181024.11+332411.1 & 1RXS J181024.3+332402 & 2.8808 & 10.26 & 9.64 & 8.31 & PW Her & RS CVn/Algol & 2.8809 & Qian et al. 2003  \\
  & J181132.84+235512.6 & 1RXS J181132.2+235516 & 8.4645 & 10.79 & 11.75 & 10.05 & V836 Her & Algol & 8.4678 & Brelstaff 1991  \\
2 & J181848.08+342233.8 & 1RXS J181848.1+342237 & 8.5030 & 12.40 & 13.22 & 11.66 & ROTSE1 J181848.06+342234.6 & Cepheid & 8.35521 & Akerlof et al. 2000  \\
  & J182538.72+181740.1 & 1RXS J182538.4+181741 & 8.8007 & 9.74 & 9.24 & 7.94 & AW Her & RS CVn/Algol & 8.80076 & Hall \& Kreiner 1980  \\
  & J182656.97+174402.4 & 1RXS J182656.6+174400 & 31.8786 & 9.43 & 8.18 & 6.73 & V992 Her & semi-reg. pulsator & 31.786 & Koen \& Eyer 2002  \\
  & J182913.01+064713.7 & 1RXS J182912.6+064717 & 0.3753 & 9.70 & 9.95 & 9.06 & HD170451 & W UMa  & 0.375296 & Koppelman et al. 2002  \\
2 & J182931.47+223424.4 & 1RXS J182931.3+223426 & 7.6650 & 11.48 & 12.74 & 10.74 & V1011 Her & Algol & ? &  \\
  & J183053.08+485848.6 & 2RXP J183052.8+485840 & 0.3476 & 11.64 & 12.11 & 11.15 & ROTSE1 J183052.99+485848.8 & W UMa  & 0.347625 & Akerlof et al. 2000  \\
  & J184633.09+485444.8 & 1RXS J184632.2+485443 & 5.0703 & 11.80 & 12.82 & 11.59 & ROTSE1 J184633.30+485435.3 & Cepheid & 5.09123 & Akerlof et al. 2000  \\
  & J204300.17+264833.7 & 1RXS J204300.0+264845 & 2.1547 & 11.24 & 10.25 & 8.63 & V401 Vul & BY Dra & 1.9252 & Koen \& Eyer 2002  \\
  & J215443.38+143327.8 & 1RXS J215443.6+143325 & 14.7539 & 9.21 & 8.86 & 7.67 & DF Peg & Algol & 14.6987 & Brancewicz \& Dworak 1980  \\
  & J225154.86+314457.5 & 1RXS J225153.6+314513 & 1.6404 & 11.12 & 13.22 & 11.13 & GT Peg & BY Dra/UV Cet  & 1.641 & Alekseev 1998  \\
  & J231653.34+254310.1 & 1RXS J231653.6+254312 & 11.6361 & 9.54 & 9.12 & 7.55 & EZ Peg & RS CVn & 11.766 & Koen \& Eyer 2002  \\ \hline
\end{longtable}
\end{landscape}
\normalsize

\tiny
\begin{longtable}{lllrrrrlll}
\caption{SuperWASP objects showing periodic variability, coincident 
with {\em ROSAT} sources, newly identified as periodic variable stars.}\\
\hline \hline
N & SuperWASP ID & {\em ROSAT} ID & Period/d  & SW V & B1 & R1 & Name & Type & Sp \\ \hline
\endfirsthead
\caption{continued.}\\ \hline \hline
N & SuperWASP ID & {\em ROSAT} ID & Period/d  & SW V & B1 & R1 & Name & Type & Sp \\ \hline
\endhead
\hline
\endfoot
  & J001357.58+350243.4 & 1RXS J001357.7+350233 & 2.5730 & 11.66 & 12.67 & 11.35 & & &  \\ 
  & J001614.18+195144.2 & 1RXS J001614.0+195142 & 4.7901 & 11.01 & 13.08 & 7.88 & LHS 107 & double star & M4  \\ 
  & J001736.91+305119.2 & 2RXP J001737.2+305119 & 13.5338 & 12.49 & 12.91 & 11.66 & & &  \\ 
  & J001825.00+232434.2 & 1RXS J001825.5+232432 & 1.5371 & 10.16 & 10.68 & 9.68 & SAO 73880 & & G0  \\ 
  & J002001.09+275954.0 & 1RXS J002001.1+275949 & 4.6817 & 10.24 & 10.96 & 9.71 & SAO73896 & & G5  \\ 
  & J002122.99+334237.1 & 1RXS J002123.2+334236 & 8.3490 & 11.01 & 11.80 & 9.84 & & &  \\ 
  & J002334.66+201428.6 & 1RXS J002334.9+201418 & 7.9165 & 10.84 & 12.54 & 10.04 & StKM 1-34 & & K5  \\ 
  & J003408.48+252349.7 & 1RXS J003408.7+252342 & 3.1555 & 11.11 & 13.00 & 10.37 & BPM 84322 & & K5  \\ 
  & J004836.93+320859.2 & 2RXP J004836.7+320857 & 0.3054 & 12.64 & 13.23 & 11.92 & & &  \\ 
2 & J005033.00+244902.0 & 1RXS J005033.3+244901 & 1.6968 & 11.48 & 14.01 & 14.96  & LP 350-19/GJ 3060A & flare star & \\
  & J005101.78+200824.4 & 1RXS J005101.5+200827 & 0.7966 & 11.77 & 12.75 & 11.04 & & &  \\ 
  & J005515.01+301515.6 & 2RXP J005514.6+301517 & 4.0674 & 13.31 & 15.69 & 12.39 & & &  \\ 
  & J005601.22+303825.9 & 2RXP J005601.2+303825 & 3.7947 & 12.88 & 12.44 & 10.55 & & &  \\ 
  & J010705.51+190908.3 & 1RXS J010703.8+190858 & 1.3754 & 10.15 & 11.12 & 9.54 & & &  \\ 
  & J011235.03+170355.7 & 1RXS J011235.6+170401 & 1.0362 & 13.55 & 15.30 & 12.92 & & &  \\ 
  & J012139.75+253642.4 & 1RXS J012139.6+253634 & 0.5649 & 9.74 & 10.13 & 9.33 & SAO 74655 & & F8  \\ 
  & J012215.44+202130.4 & 1RXS J012215.4+202139 & 10.2318 & 9.66 & 10.39 & 9.01 & & &  \\ 
  & J012457.96+255702.4 & 1RXS J012458.8+255703 & 3.0420 & 10.75 & 11.40 & 10.29 & & &  \\ 
  & J012757.37+185924.9 & 2RXP J012757.8+185931 & 0.7859 & 9.36 & 9.97 & 9.03 & BD+18 193 & & F8  \\ 
  & J013147.20+384803.2 & 1RXS J013146.8+384757 & 8.7733 & 11.49 & 12.69 & 10.99 & & &  \\ 
  & J013514.32+211622.4 & 2RXP J013514.3+211615 & 1.8710 & 10.97 & 11.58 & 10.49 & & &  \\ 
  & J013612.29+304902.5 & 2RXP J013612.1+304901 & 0.6647 & 13.04 & 14.14 & 12.00 & & &  \\ 
  & J013626.24+404343.8 & 1RXS J013625.8+404352 & 0.4357 & 12.59 & 14.07 & 11.41 & & &  \\ 
2 & J013627.81+250835.5 & 1RXS J013628.0+250835 & 3.9346 & 11.02 & 12.53 & 10.48 & BD+24 238&emission line star&K2e\\
  & J013723.23+265712.1 & 1RXS J013723.4+265709 & 1.0852 & 10.95 & 13.02 & 9.96 & StKM 1-174 & & K5  \\ 
  & J013727.14+390008.3 & 1RXS J013727.7+390000 & 1.0616 & 10.38 & 11.04 & 9.93 & & &  \\ 
  & J014028.77+421200.8 & 1RXS J014028.6+421159 & 1.0605 & 10.61 & 11.46 & 9.81 & BD+41 324 & & K0V  \\ 
  & J014453.62+282457.6 & 1RXS J014454.5+282440 & 1.1990 & 11.04 & 12.45 & 10.22 & & &  \\ 
2 & J014633.48+331711.5 & 1RXS J014631.3+331715 & 4.9393 & 10.23 & 10.89 & 9.62 & & &  \\ 
  & J015203.00+374808.7 & 1RXS J015201.5+374804 & 1.9530 & 11.53 & 11.90 & 10.96 & & &  \\ 
  & J015548.02+242606.0 & 1RXS J015548.2+242620 & 3.2541 & 10.58 & 11.81 & 10.31 & & &  \\ 
  & J015935.57+234852.6 & 1RXS J015935.7+234848 & 5.5569 & 12.54 & 13.37 & 11.38 & & &  \\ 
  & J021208.77+270818.2 & 1RXS J021208.9+270817 & 0.3182 & 10.06 & 10.56 & 9.73 & & &  \\ 
  & J022050.85+332047.6 & 1RXS J022050.7+332049 & 0.1926 & 12.91 & 14.47 & 11.80 & & &  \\ 
  & J022133.29+340445.4 & 1RXS J022132.9+340449 & 3.6159 & 9.89 & 10.54 & 9.26 & BD+33 411 & & G0  \\ 
  & J022452.45+422653.7 & 2RXP J022452.0+422654 & 2.0452 & 15.65 & 15.52 & 14.16 & & &  \\ 
2 & J022729.25+305824.6 & 1RXS J022728.4+305828 & 13.6928 & 9.96 & 11.67 & 9.32 & AG Tri & BY Dra  &  \\ 
  & J022734.78+285830.2 & 1RXS J022733.2+285834 & 4.3104 & 9.89 & 10.51 & 9.40 & SAO 75375 & & G0  \\ 
  & J022936.37+342343.1 & 1RXS J022936.5+342334 & 2.7458 & 12.20 & 13.65 & 11.76 & & &  \\ 
  & J023503.81+313922.1 & 1RXS J023504.2+313927 & 1.2771 & 10.49 & 11.12 & 9.89 & SAO 55677 & & G0  \\ 
  & J025020.65+372902.0 & 1RXS J025020.0+372909 & 1.7330 & 12.05 & 13.56 & 10.87 & & PMS star  & K4V  \\ 
  & J025217.58+361648.1 & 1RXS J025216.9+361658 & 7.9809 & 11.18 & 12.09 & 9.89 & & PMS star  & K2IV  \\ 
  & J025742.76+235744.5 & 1RXS J025740.5+235755 & 3.3348 & 10.11 & 9.37 & 8.19 & AP Ari & PMS star  & K0  \\ 
  & J025752.74+415135.1 & 1RXS J025755.5+415159 & 0.4854 & 12.06 & 13.16 & 10.52 & & &  \\ 
  & J025828.76+294753.7 & 1RXS J025828.0+294805 & 0.6741 & 11.61 & 12.51 & 10.70 & & PMS star  & K0IV  \\ 
  & J025953.10+380148.1 & 1RXS J025952.4+380149 & 0.3874 & 11.09 & 11.73 & 10.34 & & PMS star  & K0IV  \\ 
  & J030335.64+362631.2 & 1RXS J030335.4+362635 & 8.2516 & 8.25 & 7.35 & 6.55 & SAO 56117 & double star & F8  \\ 
  & J030349.82+250234.0 & 1RXS J030349.4+250241 & 2.6754 & 11.89 & 13.62 & 10.94 & & &  \\ 
  & J030405.14+300309.6 & 1RXS J030405.0+300312 & 1.8070 & 11.10 & 11.71 & 10.94 & & PMS star  & K0V  \\ 
2 & J031531.89+260449.9 & 1RXS J031530.7+260451 & 9.3197 & 12.31 & 14.40 & 11.48 & & &  \\ 
  & J032231.55+285319.8 & 1RXS J032231.4+285330 & 1.6648 & 10.76 & 11.48 & 10.43 & & &  \\ 
  & J032714.41+272309.1 & 1RXS J032714.9+272318 & 10.0933 & 11.47 & 12.98 & 8.64 & LP 300-3 & &  \\ 
  & J033025.95+310217.9 & 2RXP J033025.5+310215 & 2.2308 & 14.00 & 15.71 & 13.34 & & &  \\ 
  & J033040.80+313658.1 & 2RXP J033040.9+313657 & 1.6559 & 11.88 & 13.33 & 11.64 & BSD 47-661 & & K3V  \\ 
  & J033529.90+311337.4 & 2RXP J033529.9+311338 & 1.8366 & 9.21 & 9.55 & 8.59 & SAO 56567 & PMS star  & G0  \\ 
  & J034057.81+311805.8 & 2RXP J034058.1+311802 & 4.2419 & 11.28 & 12.06 & 11.10 & & & G1V  \\ 
  & J034145.61+271856.5 & 1RXS J034145.2+271855 & 2.6380 & 11.96 & 13.05 & 10.98 & Wolf 1260 & PMS star  & K2IV  \\ 
2 & J034220.86+291440.9 & 1RXS J034221.6+291443 & 2.3271 & 10.67 & 12.49 & 11.14 & & &  \\ 
  & J034348.34+250015.7 & 2RXP J034348.1+250008 & 0.4749 & 11.75 & 11.99 & 11.69 & NSV 15748 & &  \\ 
  & J034433.95+395948.0 & 1RXS J034432.1+395937 & 0.2888 & 12.31 & 14.53 & 13.07 & & PMS star  & K4V  \\ 
2 & J034450.16+321906.7 & 2RXP J034450.5+321909 & 6.0371 & 10.96 & 16.74 & 8.10 & Dust Ball & PMS star & A0  \\ 
  & J034557.94+273335.2 & 1RXS J034557.2+273331 & 6.6237 & 10.94 & 12.21 & 10.23 & HD 282932 & & K5  \\ 
  & J034630.44+330234.5 & 2RXP J034630.6+330238 & 1.1202 & 10.72 & 11.60 & 9.83 & HD 278996 & & F  \\ 
  & J034906.11+234653.0 & 2RXP J034906.2+234652 & 0.3082 & 11.08 & 11.05 & 12.00 & & & G9  \\ 
  & J034936.53+241745.8 & 2RXP J034936.3+241752 & 5.9466 & 11.97 & 12.49 & 10.88 & V468 Tau & UV Cet  & M3.7  \\ 
  & J034942.26+242746.8 & 2RXP J034942.2+242737 & 7.7479 & 12.27 & 13.48 & 11.39 & NSV 1358 & & K3  \\ 
  & J035054.31+235005.5 & 1RXS J035055.0+235016 & 5.4490 & 11.76 & 12.44 & 10.99 & V1176 Tau & BY Dra  &  \\ 
  & J035208.32+241348.5 & 2RXP J035208.5+241343 & 1.6604 & 11.05 & 11.42 & 10.14 & HD 283063 & & G5  \\ 
  & J035331.35+263141.1 & 1RXS J035330.5+263152 & 0.2835 & 12.13 & 12.71 & 11.16 & & & G7IV  \\ 
  & J035425.23+242136.2 & 1RXS J035423.8+242146 & 2.2386 & 11.38 & 10.62 & 9.60 & HD 283167 & & G5IV  \\ 
  & J035525.61+313047.9 & 2RXP J035525.1+313048 & 0.4856 & 12.98 & 14.29 & 12.11 & & &  \\ 
  & J040005.79+394137.2 & 1RXS J040005.4+394133 & 5.6162 & 12.09 & 13.21 & 11.08 & & &  \\ 
  & J040031.06+193520.8 & 1RXS J040030.7+193521 & 1.1677 & 10.44 & 11.51 & 9.74 & HD 285281 & PMS star  & K0  \\ 
  & J040105.23+343902.9 & 2RXP J040105.2+343906 & 18.1451 & 10.73 & 11.77 & 10.23 & HD 279311 & & K0  \\ 
2 & J040519.59+200925.5 & 1RXS J040518.6+200919 & 1.4406 & 10.55 & 11.27 & 9.57 & & &  \\ 
  & J040753.30+335605.0 & 1RXS J040753.0+335555 & 3.8678 & 11.43 & 12.64 & 10.65 & & & K3  \\ 
  & J040754.31+352749.2 & 1RXS J040753.8+352730 & 9.4797 & 10.45 & 11.54 & 9.75 & HD 279444 & & G0  \\ 
  & J041327.27+281625.0 & 2RXP J041327.5+281623 & 0.8637 & 12.69 & 15.61 & 11.90 & V1096 Tau & BY Dra  & M0V  \\ 
2 & J041430.63+285129.8 & 2RXP J041431.0+285124 & 3.6916 & 11.15 & 14.17 & 11.26 & V1097 Tau & emission line star &\\
  & J041738.93+283300.7 & 2RXP J041739.1+283255 & 1.4127 & 13.26 & 15.57 & 13.12 & LkCa5 & PMS star  & M2V  \\ 
  & J041810.78+231704.7 & 1RXS J041811.1+231700 & 1.8797 & 9.72 & 10.51 & 8.93 & HD 284303 & & K0  \\ 
2 & J041831.11+281629.1 & 2RXP J041831.4+281636 & 5.7593 & 13.29 & 15.98 & 13.76 & DD Tau & PMS star  & K6V  \\ 
  & J041846.99+282008.2 & 2RXP J041847.0+282012 & 1.5483 & 12.00 & 14.31 & 11.21 & V1023 Tau & PMS star  & K7  \\ 
  & J041946.58+231748.4 & 1RXS J041946.0+231750 & 2.2176 & 10.82 & 11.68 & 10.25 & HD 284296 & & G5  \\ 
  & J041953.70+300953.7 & 1RXS J041953.5+300949 & 10.2706 & 11.26 & 12.83 & 10.13 & HD 281912 & & K0  \\ 
  & J042347.60+294038.2 & 1RXS J042346.8+294033 & 1.3836 & 11.51 & 12.73 & 11.52 & & & K2  \\ 
  & J042521.02+254257.0 & 2RXP J042521.4+254252 & 7.6157 & 14.15 & 16.39 & 13.07 & TAP34 & PMS star &  \\ 
  & J042637.39+384502.3 & 1RXS J042638.5+384458 & 2.0596 & 11.04 & 11.72 & 10.24 & HD 279788 & PMS star  & G5V  \\ 
  & J042937.53+232033.2 & 1RXS J042937.8+232032 & 4.5854 & 10.52 & 11.56 & 9.65 & HD 284475 & double star & K2  \\ 
  & J043049.18+211410.6 & 1RXS J043049.5+211356 & 0.7397 & 10.48 & 11.10 & 10.05 & HD 284503 & PMS star  & G8  \\ 
  & J043114.43+271017.9 & 1RXS J043114.5+271021 & 5.9087 & 12.62 & 14.33 & 11.75 & JH 56 & PMS star & M1  \\ 
2 & J043230.55+241957.6 & 2RXP J043230.9+242002 & 6.6910 & 13.49 & 18.05 & 14.17 & FY Tau & PMS star  &  \\ 
  & J043547.33+225021.4 & 2RXP J043547.3+225027 & 2.4546 & 12.02 & 14.15 & 11.18 & HQ Tau & PMS star  &  \\ 
5 & J043554.15+225413.5 & 1RXS J043553.6+225410 & 1.2072 & 11.08 & 12.60 & 10.14 & V1025 Tau & PMS star  &  \\ 
  & J043619.09+254259.0 & 1RXS J043618.5+254255 & 3.3353 & 11.89 & 13.04 & 10.81 & V1115 Tau & BY Dra  & M0V  \\ 
2 & J043925.45+333244.6 & 1RXS J043926.1+333218 & 2.4180 & 11.75 & 13.12 & 10.45 & & &  \\ 
2 & J043931.00+340744.5 & 1RXS J043931.0+340737 & 0.7333 & 9.94 & 9.22 & 8.04 & HD 282346 & double star & K2  \\ 
  & J044356.93+372303.4 & 1RXS J044357.5+372303 & 4.2878 & 12.98 & 15.35 & 12.75 & & &  \\ 
  & J044721.02+280853.1 & 1RXS J044720.6+280903 & 0.6232 & 13.04 & 15.72 & 13.20 & & &  \\ 
  & J045222.04+400634.8 & 1RXS J045222.6+400633 & 6.2542 & 12.47 & 14.03 & 12.38 & & &  \\ 
  & J045450.62+320411.8 & 1RXS J045450.5+320406 & 0.2801 & 11.43 & 12.22 & 10.59 & HD 282589 & & G0  \\ 
2 & J045730.52+223513.5 & 1RXS J045730.8+223458 & 4.6635 & 10.79 & 11.65 & 10.66 & HD 284988 & & G  \\ 
  & J045808.96+433301.1 & 1RXS J045809.3+433257 & 0.4411 & 11.98 & 12.35 & 11.12 & & &  \\ 
  & J050006.91+240834.3 & 2RXP J050006.9+240833 & 8.6789 & 13.18 & 15.79 & 13.70 & & &  \\ 
  & J050147.63+380541.9 & 1RXS J050147.7+380545 & 2.8542 & 11.01 & 11.37 & 10.54 & V526 Aur & BY Dra  & G0  \\ 
  & J050206.19+311102.2 & 1RXS J050205.8+311111 & 4.2336 & 11.07 & 11.82 & 10.59 & HD 282718 & & G5  \\ 
  & J050206.88+242739.8 & 2RXP J050206.8+242741 & 0.1394 & 14.14 & 14.87 & 12.97 & & &  \\ 
  & J050559.66+280716.8 & 1RXS J050559.4+280717 & 0.5239 & 10.18 & 9.59 & 8.60 & HD 284065 & PMS star & G0/B4  \\ 
  & J051022.34+312640.1 & 1RXS J051023.1+312648 & 2.2013 & 11.30 & 12.10 & 11.01 & & PMS star  & K2IV  \\ 
  & J051043.42+302042.7 & 1RXS J051043.5+302044 & 1.8082 & 12.01 & 13.39 & 10.39 & & &  \\ 
  & J051740.27+335355.4 & 2RXP J051740.6+335357 & 0.4549 & 12.09 & 12.93 & 11.30 & & &  \\ 
  & J051908.49+340537.5 & 1RXS J051908.2+340529 & 2.1662 & 12.30 & 13.52 & 12.48 & & PMS star  & K2IV  \\ 
  & J052037.06+244713.5 & 1RXS J052036.6+244731 & 0.6931 & 11.82 & 12.48 & 10.88 & & &  \\ 
2 & J052146.83+240044.4 & 1RXS J052146.7+240036 & 3.4547 & 10.85 & 11.70 & 9.77 & & PMS star  & G7IV  \\ 
  & J052210.33+243208.8 & 1RXS J052210.2+243200 & 2.5206 & 10.94 & 11.94 & 10.36 & HD 242903 & & G0  \\ 
  & J052247.17+243731.1 & 1RXS J052248.0+243731 & 1.3385 & 11.95 & 13.00 & 10.77 & & &  \\ 
2 & J052354.67+253047.0 & 1RXS J052355.2+253052 & 7.7241 & 12.32 & 13.49 & 11.45 & & PMS star  & G4V  \\ 
2 & J052638.31+223152.5 & 1RXS J052638.7+223151 & 4.6765 & 11.61 & 13.16 & 10.81 & & PMS star x2 & K2IV/K0IV \\ 
  & J052705.88+213525.5 & 1RXS J052706.4+213525 & 0.5477 & 11.26 & 12.34 & 10.45 & & PMS star  & G7IV  \\ 
2 & J052942.46+233411.0 & 1RXS J052943.1+233412 & 1.6347 & 13.05 & 14.45 & 12.41 & & &  \\ 
  & J053020.91+414914.0 & 1RXS J053021.2+414914 & 0.7299 & 11.48 & 11.61 & 10.72 & & &  \\ 
  & J053104.38+231234.7 & 1RXS J053103.9+231232 & 0.9426 & 9.11 & 8.87 & 7.82 & SAO 77222 & PMS star  & G0  \\ 
  & J053505.64+394631.7 & 1RXS J053506.4+394644 & 4.5055 & 10.83 & 11.48 & 9.74 & & PMS star  & K0V  \\ 
  & J053807.15+422029.1 & 1RXS J053807.4+422039 & 3.3000 & 12.12 & 12.91 & 10.99 & & &  \\ 
  & J053858.03+244257.1 & 1RXS J053858.2+244320 & 4.3542 & 12.43 & 14.35 & 11.57 & & &  \\ 
  & J060358.42+222833.4 & 2RXP J060358.3+222829 & 0.3188 & 11.80 & 12.38 & 10.87 & & & K0III  \\ 
  & J060951.04+322948.3 & 1RXS J060951.5+323011 & 7.8858 & 10.39 & 11.51 & 9.46 & & &  \\ 
  & J061015.84+211956.3 & 1RXS J061016.0+212006 & 4.0814 & 11.34 & 12.43 & 10.50 & G104-22 &  & K2  \\ 
  & J061128.82+202149.5 & 1RXS J061128.8+202151 & 8.6016 & 10.18 & 11.14 & 9.51 & & &  \\ 
5 & J061510.08+252248.1 & 1RXS J061509.9+252250 & 0.9610 & 11.58 & 12.05 & 10.82 & & &  \\ 
  & J062403.09+225336.2 & 1RXS J062402.8+225334 & 0.7696 & 10.61 & 11.51 & 9.79 & & &  \\ 
  & J065341.90+424219.4 & 1RXS J065341.6+424217 & 3.5416 & 11.28 & 12.38 & 10.90 & & & K  \\ 
  & J065948.41+274158.7 & 1RXS J065948.1+274200 & 0.1965 & 10.15 & 10.71 & 9.39 & & &  \\ 
2 & J070818.57+310508.1 & 1RXS J070818.1+310520 & 1.1200 & 11.14 & 11.59 & 10.87 & & &  \\ 
  & J071341.07+273103.7 & 1RXS J071340.5+273109 & 4.2860 & 11.50 & 12.37 & 10.61 & & &  \\ 
  & J072343.59+202458.6 & 1RXS J072343.6+202500 & 2.7835 & 10.07 & 11.31 & 9.43 & BD+20 1790 &  & K3  \\ 
2 & J114608.99+400151.9 & 1RXS J114608.2+400156 & 1.3672 & 12.15 & 13.89 & 11.68 & & &  \\ 
  & J114823.63+350421.2 & 1RXS J114824.5+350435 & 14.7895 & 11.64 & 12.38 & 10.97 & & &  \\ 
  & J114839.23+231138.6 & 1RXS J114839.6+231136 & 4.7245 & 10.44 & 11.19 & 9.60 & & &  \\ 
  & J114903.63+380031.4 & 1RXS J114904.2+380038 & 8.7540 & 11.00 & 11.82 & 10.06 & BD+38 2281 & & G7III  \\ 
  & J121911.62+291201.3 & 2RXP J121912.2+291150 & 2.1677 & 10.28 & 10.58 & 9.88 & BD+30 2254 & & F6  \\ 
  & J122224.73+334614.5 & 2RXP J122224.5+334616 & 0.1794 & 12.41 & 12.33 & 12.41 & & & F8  \\ 
  & J122354.88+224546.0 & 2RXP J122354.8+224547 & 1.8163 & 12.12 & 13.29 & 11.38 & & &  \\ 
  & J122557.80+334651.1 & 2RXP J122557.6+334650 & 1.3636 & 11.97 & 12.59 & 10.85 & & & G0V  \\ 
  & J122751.20+333843.4 & 2RXP J122751.2+333848 & 12.5865 & 11.90 & 13.57 & 11.20 & & &  \\ 
  & J123013.72+215810.5 & 1RXS J123013.7+215812 & 1.1489 & 10.23 & 9.76 & 9.08 & BD+22 2477 & & G0  \\ 
  & J123036.52+343027.9 & 1RXS J123034.0+343033 & 0.2491 & 9.42 & 9.53 & 8.42 & & & K4V  \\ 
  & J123240.68+234804.8 & 2RXP J123240.8+234806 & 7.5566 & 10.19 & 10.99 & 9.83 & & &  \\ 
  & J123703.97+374456.5 & 2RXP J123704.4+374502 & 0.7941 & 13.61 & 14.75 & 12.86 & & &  \\ 
  & J124408.46+402329.8 & 1RXS J124409.3+402328 & 2.6454 & 11.28 & 11.96 & 11.13 & BD+41 2326 & & G5  \\ 
  & J124705.32+362844.1 & 1RXS J124706.0+362854 & 7.0385 & 11.75 & 12.83 & 11.83 & & & G8III  \\ 
  & J125147.23+223239.4 & 1RXS J125146.8+223240 & 3.4046 & 10.48 & 9.70 & 7.99 & BD+23 2508 & & M0  \\ 
  & J125532.88+301110.7 & 1RXS J125532.4+301108 & 2.5903 & 11.23 & 12.28 & 10.37 & & & K8V  \\ 
  & J125535.36+270349.3 & 1RXS J125534.6+270355 & 4.4552 & 10.97 & 12.14 & 10.68 & BD+27 2192 & & K0IV  \\ 
  & J130345.95+283720.6 & 1RXS J130346.2+283729 & 3.2035 & 10.76 & 11.59 & 10.25 & & &  \\ 
  & J131308.01+294052.4 & 2RXP J131309.0+294059 & 9.1128 & 11.08 & 11.80 & 10.63 & & & G7V  \\ 
  & J132117.60+210124.9 & 1RXS J132117.1+210116 & 1.3456 & 11.82 & 12.71 & 11.27 & & &  \\ 
  & J132426.35+303314.2 & 2RXP J132426.4+303316 & 3.3479 & 9.83 & 9.87 & 9.52 & HD 116635 & & F2  \\ 
  & J132712.10+455826.4 & 1RXS J132713.0+455826 & 2.1838 & 11.38 & 12.62 & 10.50 & & &  \\ 
  & J132745.35+474546.7 & 1RXS J132747.0+474549 & 5.7176 & 8.89 & 10.64 & 8.50 & HD 117174 & double star & K0/K0 \\ 
  & J132837.25+353311.6 & 2RXP J132838.0+353303 & 0.1670 & 12.09 & 12.82 & 11.43 & & &  \\ 
  & J132931.19+293616.2 & 2RXP J132931.3+293622 & 7.9604 & 10.38 & 11.22 & 9.81 & & &  \\ 
  & J133240.48+284758.1 & 1RXS J133240.3+284747 & 2.4849 & 10.98 & 11.63 & 10.50 & & &  \\ 
3 & J133319.04+230050.9 & 1RXS J133318.7+230110 & 0.3525 & 9.41 & 10.62 & 9.37 & SAO 82877 & double star & G5/G5 \\ 
  & J133538.39+491406.1 & 2RXP J133538.3+491406 & 0.2938 & 11.02 & 11.70 & 10.45 & & &  \\ 
  & J133924.66+400904.8 & 2RXP J133923.4+400904 & 16.2579 & 11.04 & 11.94 & 10.22 & & &  \\ 
  & J134328.21+391132.7 & 1RXS J134327.7+391131 & 3.0543 & 11.36 & 12.06 & 11.12 & & &  \\ 
  & J134330.73+332951.2 & 1RXS J134330.5+332946 & 5.9383 & 11.25 & 12.03 & 11.23 & & &  \\ 
  & J135418.84+404542.3 & 1RXS J135418.6+404533 & 0.3454 & 9.25 & 9.94 & 8.95 & SAO 44739 & &  G0 \\	
  & J142004.68+390301.5 & 1RXS J142004.8+390309 & 0.3693 & 11.79 & 14.26 & 11.17 & GJ3842 & flare star & M2.5  \\ 
  & J142056.95+345952.4 & 1RXS J142057.2+345958 & 2.0295 & 9.94 & 11.34 & 9.83 & BD+35 2550 & double star & K2  \\ 
  & J142643.20+315216.0 & 1RXS J142643.5+315221 & 21.3193 & 10.25 & 11.48 & 9.64 & & &  \\ 
  & J142656.60+233652.8 & 2RXP J142656.3+233657 & 6.4131 & 10.90 & 11.37 & 10.37 & & &  \\ 
  & J142902.52+335038.8 & 1RXS J142903.2+335033 & 3.8028 & 9.32 & 10.03 & 8.80 & SAO 64168 & & G  \\ 
  & J143729.47+412835.1 & 1RXS J143729.6+412842 & 2.0896 & 13.15 & 15.50 & 12.83 & & & M  \\ 
  & J143854.57+330019.9 & 1RXS J143854.8+330022 & 5.0682 & 11.12 & 12.23 & 10.84 & & &  \\ 
  & J144145.29+423121.1 & 1RXS J144145.0+423124 & 0.5940 & 11.39 & 12.29 & 11.30 & & & \\	
  & J144831.00+350318.0 & 1RXS J144831.7+350329 & 3.4828 & 10.84 & 11.42 & 10.15 & & &  \\ 
  & J144952.59+420626.7 & 1RXS J144952.3+420615 & 13.7134 & 11.17 & 12.20 & 10.27 & & &  \\ 
  & J150018.89+335206.6 & 1RXS J150019.4+335152 & 2.4429 & 11.81 & 12.52 & 11.02 & & &  \\ 
  & J152152.93+205839.9 & 1RXS J152153.0+205830 & 3.3829 & 9.78 & 9.29 & 6.61 & OT Ser & BY Dra  & M9  \\ 
  & J153352.71+311801.4 & 2RXP J153352.7+311802 & 1.5262 & 12.14 & 13.29 & 11.31 & & &  \\ 
  & J153440.64+265442.7 & 2RXP J153440.6+265442 & 0.3322 & 9.04 & 9.49 & 8.47 & SAO 83894 & & G0  \\ 
  & J153511.33+384358.7 & 1RXS J153511.5+384344 & 2.1942 & 12.09 & 12.99 & 11.65 &  \\ 
  & J153633.39+271029.2 & 2RXP J153633.5+271037 & 1.3393 & 10.00 & 9.75 & 9.37 & SAO 83906 & & A0  \\ 
  & J153650.34+373449.4 & 1RXS J153651.3+373446 & 4.3950 & 11.30 & 10.26 & 8.40 & G 179-40 &  & M2  \\ 
  & J153704.08+374827.8 & 1RXS J153704.2+374830 & 1.2452 & 13.30 & 15.26 & 12.73 & & & K  \\ 
  & J154058.90+402700.2 & 1RXS J154058.3+402657 & 1.9775 & 10.50 & 11.59 & 10.27 & & &  \\ 
  & J154535.07+420506.6 & 1RXS J154532.3+420500 & 5.7211 & 13.67 & 15.97 & 13.77 & & &  \\ 
  & J155307.49+202838.8 & 2RXP J155307.6+202846 & 9.2291 & 12.03 & 13.33 & 11.19 & & &  \\ 
  & J155431.35+295652.1 & 1RXS J155430.9+295637 & 0.2328 & 11.46 & 12.46 & 11.35 & & &  \\ 
  & J155842.13+323046.0 & 1RXS J155842.6+323047 & 6.2612 & 11.86 & 13.71 & 10.74 & & &  \\ 
3 & J155850.50+272327.1 & 2RXP J155851.6+272333 & 34.2257 & 13.07 & 14.18 & 11.90 & & &  \\ 
  & J160248.22+252038.2 & 1RXS J160248.3+252031 & 0.4955 & 10.71 & 11.55 & 10.16 & & &  \\ 
  & J160351.74+423654.4 & 2RXP J160351.6+423650 & 0.7880 & 12.89 & 14.33 & 11.68 & & &  \\ 
  & J160713.97+340136.0 & 1RXS J160714.4+340123 & 0.7418 & 10.91 & 12.11 & 10.46 & & &  \\ 
  & J161213.07+341416.2 & 2RXP J161212.6+341411 & 0.6031 & 13.92 & 15.29 & 13.09 & & &  \\ 
  & J162013.71+243611.0 & 1RXS J162013.2+243606 & 18.5948 & 9.74 & 10.76 & 8.95 & SAO 84309 & & K0  \\ 
  & J162201.18+225021.6 & 1RXS J162200.9+225009 & 1.3728 & 11.83 & 13.70 & 11.10 & & &  \\ 
  & J162255.30+224604.1 & 1RXS J162255.0+224559 & 1.7387 & 10.86 & 11.59 & 10.18 & & & K  \\ 
  & J162641.33+335041.8 & 1RXS J162640.6+335033 & 23.1808 & 9.40 & 9.18 & 7.71 & SAO 65302 & & G0  \\ 
  & J162946.59+281038.0 & 1RXS J162946.1+281034 & 1.4639 & 10.68 & 11.28 & 10.01 & & &  \\ 
  & J163052.87+241224.3 & 2RXP J163052.5+241219 & 53.4764 & 10.80 & 11.79 & 10.00 & & &  \\ 
  & J163420.90+424433.4 & 2RXP J163420.4+424426 & 0.3636 & 10.86 & 11.42 & 10.54 & & &  \\ 
2 & J163527.45+350057.7 & 1RXS J163527.5+350046 & 0.9166 & 12.23 & 14.35 & 11.80 & GJ 3966 & flare star & M4  \\ 
  & J163739.45+221112.8 & 1RXS J163739.5+221104 & 2.0912 & 10.84 & 11.78 & 10.55 & & &  \\ 
  & J163741.35+291950.4 & 1RXS J163741.2+291946 & 0.8246 & 11.43 & 12.70 & 10.82 & & &  \\ 
  & J164159.14+363818.5 & 2RXP J164159.7+363819 & 12.2277 & 12.20 & 13.04 & 11.34 & & &  \\ 
3 & J164732.08+251938.5 & 1RXS J164732.3+251931 & 0.9048 & 11.50 & 13.27 & 10.78 & & &  \\ 
  & J164942.92+222003.7 & 1RXS J164943.4+222009 & 22.9494 & 10.34 & 11.83 & 9.37 & & &  \\ 
  & J164956.83+325235.6 & 1RXS J164955.5+325232 & 1.0396 & 11.98 & 13.61 & 11.23 & & &  \\ 
  & J164959.95+412225.9 & 1RXS J165000.2+412217 & 1.8137 & 11.24 & 12.43 & 10.50 & AG+41 1420 & & K0  \\ 
  & J165025.84+272817.2 & 1RXS J165027.1+272812 & 2.2815 & 11.32 & 11.95 & 11.12 & & &  \\ 
  & J165107.29+283601.0 & 1RXS J165107.2+283601 & 16.1973 & 11.44 & 13.83 & 10.87 & StKM 1-1407 & & K7  \\ 
2 & J165211.93+202138.7 & 1RXS J165212.4+202148 & 3.6987 & 12.43 & 14.67 & 11.25 & & &  \\ 
  & J165445.06+423227.5 & 1RXS J165445.3+423237 & 0.6912 & 13.02 & 14.57 & 12.82 & & &  \\ 
  & J165820.66+333353.2 & 1RXS J165819.2+333411 & 4.5302 & 9.81 & 10.52 & 9.35 & & &  \\ 
  & J165909.65+205816.4 & 1RXS J165909.5+205807 & 4.1037 & 11.90 & 14.26 & 10.89 & & &  \\ 
  & J165921.89+342822.4 & 2RXP J165921.7+342822 & 1.5636 & 10.79 & 11.17 & 10.29 & & &  \\ 
  & J170033.82+200134.1 & 1RXS J170032.9+200137 & 4.2298 & 9.95 & 11.37 & 9.56 & SAO 84750 & double star & G5  \\ 
  & J170303.06+320325.8 & 1RXS J170303.1+320320 & 2.7718 & 11.48 & 12.07 & 10.87 & BD+32 2836 & & F8  \\ 
  & J170313.51+245321.0 & 1RXS J170313.2+245332 & 5.8417 & 10.14 & 10.67 & 9.50 & & &  \\ 
  & J170352.84+321145.7 & 1RXS J170352.9+321147 & 15.4221 & 11.27 & 13.01 & 11.43 & NLTT 44114 & & M3 \\
  & J171733.62+495515.7 & 2RXP J171733.8+495513 & 2.0939 & 13.87 & 14.76 & 14.04 & & &  \\ 
  & J171800.30+212809.4 & 1RXS J171800.1+212816 & 0.7904 & 10.08 & 10.55 & 9.66 & & &  \\ 
2 & J171808.56+250612.0 & 1RXS J171807.5+250610 & 2.4103 & 10.71 & 11.97 & 10.21 & BD+25 3238 & & M0  \\ 
  & J171921.10+480342.8 & 2RXP J171921.7+480338 & 5.9483 & 9.75 & 10.39 & 9.28 & SAO 46635 & & G5  \\ 
2 & J171954.21+263003.0 & 1RXS J171953.4+262958 & 19.8077 & 10.65 & 13.05 & 10.49 & V647 Her & UV Cet  & M4eV  \\ 
  & J171959.47+241205.6 & 1RXS J171959.4+241202 & 0.7129 & 13.35 & 15.65 & 13.43 & V475 Her & UV Cet  &  \\ 
  & J172011.53+495456.0 & 2RXP J172011.4+495454 & 0.9353 & 11.57 & 12.09 & 11.15 & & &  \\ 
  & J172158.32+574922.2 & 1RXS J172157.8+574913 & 6.4272 & 11.11 & 11.74 & 10.78 & & &  \\ 
  & J172228.64+365842.1 & 1RXS J172228.5+365843 & 1.2283 & 10.63 & 11.99 & 10.15 & StKM 1-1466 & & K5  \\ 
  & J172314.19+283650.3 & 1RXS J172314.4+283641 & 3.8881 & 10.85 & 11.66 & 10.62 & & &  \\ 
  & J172413.67+402617.2 & 1RXS J172413.5+402616 & 0.2890 & 11.38 & 12.27 & 10.55 & & &  \\ 
  & J172524.34+504212.2 & 2RXP J172522.4+504216 & 0.5137 & 14.65 & 16.03 & 14.80 & & &  \\ 
  & J173004.97+184339.3 & 1RXS J173004.9+184340 & 12.6078 & 10.69 & 11.37 & 10.09 & BD+18 3387 & & K0  \\ 
  & J173103.32+281506.5 & 1RXS J173103.4+281510 & 1.2653 & 10.27 & 11.07 & 9.72 & & &  \\ 
  & J173216.05+484750.1 & 1RXS J173216.5+484754 & 12.7178 & 12.39 & 13.83 & 12.48 & & &  \\ 
3 & J173335.83+204847.4 & 1RXS J173336.7+204847 & 7.0610 & 10.29 & 11.45 & 9.68 & & &  \\ 
  & J173353.14+165512.8 & 1RXS J173353.5+165515 & 0.2659 & 13.04 & 16.14 & 13.93 & & &  \\ 
  & J173636.82+151508.3 & 1RXS J173637.6+151511 & 2.6726 & 10.34 & 10.90 & 9.91 & & &  \\ 
  & J173658.21+300947.5 & 1RXS J173657.6+300943 & 5.1781 & 11.00 & 11.88 & 10.61 & & &  \\ 
  & J173659.28+485946.1 & 1RXS J173658.9+485931 & 2.6143 & 12.72 & 14.72 & 12.49 & StKM 1-1501 & & K4  \\ 
  & J173733.44+414619.9 & 1RXS J173734.1+414618 & 1.5480 & 11.16 & 12.25 & 10.86 & & &  \\ 
  & J174431.65+131257.5 & 1RXS J174432.1+131259 & 2.7015 & 11.44 & 12.77 & 10.34 & & &  \\ 
2 & J174625.25+222859.5 & 1RXS J174624.9+222851 & 3.5384 & 11.08 & 11.75 & 10.52 & & &  \\ 
  & J174705.04+332129.1 & 1RXS J174704.1+332126 & 3.2041 & 11.60 & 12.53 & 10.94 & & &  \\ 
2 & J174746.93+521340.2 & 1RXS J174745.8+521355 & 2.9532 & 11.42 & 12.26 & 11.18 & & &  \\ 
  & J174903.15+230745.5 & 2RXP J174903.8+230744 & 16.9438 & 11.73 & 12.64 & 10.55 & & &  \\ 
  & J174947.04+335059.1 & 1RXS J174947.6+335056 & 1.3479 & 10.90 & 11.49 & 10.05 & & &  \\ 
  & J174951.66+232807.3 & 2RXP J174951.9+232759 & 2.1447 & 9.95 & 10.39 & 9.53 & SAO 85451 & & F8  \\ 
  & J175133.95+414127.4 & 1RXS J175133.3+414121 & 9.3576 & 9.97 & 11.22 & 9.29 & BD+41 2912 & & K0  \\ 
  & J175141.38+281901.2 & 1RXS J175140.9+281855 & 2.6946 & 13.02 & 14.67 & 12.92 & & & K  \\ 
  & J175152.94+093751.8 & 2RXP J175153.4+093753 & 9.1586 & 10.86 & 11.89 & 10.73 & & &  \\ 
  & J175216.36+093757.5 & 2RXP J175216.1+093748 & 7.2957 & 10.02 & 10.59 & 9.36 & & &  \\ 
  & J175242.72+232728.9 & 1RXS J175242.3+232724 & 3.0911 & 10.22 & 11.28 & 9.74 & BD+23 3201 & double star &  \\ 
  & J175319.15+213029.5 & 1RXS J175318.5+213028 & 10.7768 & 10.89 & 12.31 & 10.54 & & &  \\ 
2 & J175540.63+372516.0 & 2RXP J175539.5+372516 & 3.1203 & 12.94 & 13.57 & 12.72 & & &  \\ 
2 & J175711.41+224706.3 & 1RXS J175711.2+224712 & 1.8139 & 11.74 & 12.65 & 11.19 & StKM 1-1560 & & K7  \\ 
  & J175718.89+313315.9 & 1RXS J175718.5+313314 & 0.6981 & 10.79 & 11.33 & 10.12 & & &  \\ 
  & J175734.12+584414.2 & 1RXS J175733.7+584414 & 3.2408 & 11.96 & 13.76 & 10.86 & & &  \\ 
  & J175758.92+550607.7 & 1RXS J175758.9+550608 & 0.6374 & 11.20 & 11.85 & 10.60 & & &  \\ 
2 & J175809.40+092240.8 & 1RXS J175809.3+092241 & 0.4885 & 10.90 & 11.79 & 10.38 & & &  \\ 
  & J175910.35+584259.4 & 1RXS J175910.1+584300 & 1.2314 & 11.00 & 11.68 & 10.63 & & &  \\ 
2 & J175954.36+104418.9 & 1RXS J175953.7+104402 & 5.8737 & 10.91 & 11.97 & 10.20 & & &  \\ 
  & J180029.35+510008.8 & 2RXP J180028.5+510002 & 1.1913 & 10.04 & 10.41 & 9.56 & SAO 30688 & & F8  \\ 
  & J180100.51+233945.4 & 1RXS J180100.0+233936 & 14.8666 & 10.77 & 11.72 & 10.09 & HD341448 & & K0  \\ 
  & J180207.45+183044.2 & 1RXS J180208.6+183043 & 0.5477 & 11.66 & 13.44 & 11.41 & & &  \\ 
  & J180238.80+335634.6 & 1RXS J180239.2+335639 & 7.4245 & 9.62 & 10.24 & 9.13 & SAO 66576 & & G5  \\ 
  & J180331.30+080836.3 & 2RXP J180331.2+080832 & 2.0523 & 12.07 & 12.17 & 11.45 & & &  \\ 
  & J180426.56+393047.1 & 1RXS J180426.3+393044 & 1.5451 & 11.84 & 13.25 & 10.98 & & &  \\ 
  & J180500.38+111013.9 & 2RXP J180500.8+111021 & 0.6070 & 11.02 & 11.32 & 10.57 & & &  \\ 
  & J180514.38+113148.6 & 2RXP J180514.3+113143 & 2.7255 & 13.08 & 13.89 & 11.88 & & &  \\ 
  & J180525.04+175729.7 & 2RXP J180525.1+175725 & 27.1661 & 11.45 & 12.19 & 11.06 & & &  \\ 
  & J180859.30+454910.7 & 2RXP J180859.1+454927 & 6.0582 & 14.37 & 15.40 & 14.01 & & &  \\ 
  & J181004.20+090620.8 & 2RXP J181004.1+090622 & 0.8470 & 14.11 & 15.76 & 13.55 & & &  \\ 
  & J181306.46+260151.9 & 1RXS J181306.1+260145 & 2.2838 & 12.77 & 14.93 & 12.64 & GJ 4044 & flare star & M4  \\ 
  & J181350.48+134936.6 & 2RXP J181350.4+134937 & 2.6248 & 13.68 & 15.55 & 13.24 & & &  \\ 
3 & J181538.78+381949.9 & 1RXS J181537.9+381927 & 3.1452 & 9.79 & 10.68 & 9.47 & & &  \\ 
  & J181725.16+482201.8 & 1RXS J181725.6+482202 & 16.2578 & 10.91 & 13.89 & 10.77 & & &  \\ 
  & J181938.10+364059.2 & 1RXS J181937.8+364057 & 1.0536 & 11.40 & 12.77 & 10.83 & & &  \\ 
  & J182131.56+233430.7 & 1RXS J182130.8+233434 & 1.0945 & 11.51 & 12.42 & 10.76 & HD 342009 & & K0  \\ 
  & J182247.09+443442.9 & 1RXS J182247.1+443442 & 5.7509 & 11.52 & 12.92 & 10.57 & & &  \\ 
  & J182850.26+350634.3 & 1RXS J182849.7+350637 & 2.6940 & 9.18 & 9.57 & 8.66 & SAO 67013 & & G5  \\ 
  & J182934.89+295804.5 & 1RXS J182935.0+295807 & 0.8610 & 8.92 & 8.74 & 7.86 & SAO 86121 & double star & F8/F5  \\ 
  & J183018.86+344656.4 & 1RXS J183018.1+344633 & 3.0479 & 10.78 & 11.28 & 10.12 & & &  \\ 
  & J183037.25+433553.2 & 1RXS J183037.6+433555 & 23.1356 & 11.39 & 12.62 & 11.82 & & &  \\ 
  & J183544.62+300814.7 & 1RXS J183544.4+300808 & 0.6175 & 11.06 & 11.57 & 10.29 & & &  \\ 
  & J183956.25+510534.0 & 1RXS J183956.7+510532 & 6.8528 & 9.01 & 8.49 & 7.52 & SAO31094 & double star & G5  \\ 
  & J202823.91+113110.9 & 1RXS J202823.9+113115 & 1.0209 & 10.02 & 10.40 & 9.47 & & &  \\ 
  & J202932.83+122730.8 & 1RXS J202932.3+122735 & 3.7553 & 9.95 & 10.42 & 9.40 & & &  \\ 
2 & J203553.02+100611.9 & 1RXS J203552.3+100555 & 0.9644 & 12.44 & 14.09 & 12.68 & & &  \\ 
  & J203621.99+121539.4 & 1RXS J203622.0+121519 & 12.3366 & 9.77 & 10.89 & 8.91 & & &  \\ 
2 & J203904.65+233847.2 & 1RXS J203905.4+233845 & 0.7145 & 12.28 & 13.65 & 11.35 & & &  \\ 
  & J204017.10+143035.6 & 1RXS J204018.3+143030 & 0.9378 & 9.86 & 10.12 & 9.38 & & &  \\ 
2 & J204404.76+131412.1 & 1RXS J204404.5+131413 & 2.1510 & 10.71 & 11.34 & 10.09 & & &  \\ 
  & J204853.35+122230.3 & 1RXS J204853.6+122219 & 6.2897 & 10.77 & 11.71 & 10.32 & BD+11 4390 & & G5  \\ 
  & J204922.85+064739.2 & 1RXS J204924.0+064737 & 9.1662 & 10.33 & 10.84 & 9.57 & BD+06 4654 & & F  \\ 
  & J205428.00+090606.6 & 1RXS J205428.0+090615 & 2.2374 & 11.73 & 12.80 & 10.58 & & &  \\ 
  & J210124.59+054212.8 & 1RXS J210124.1+054207 & 0.9801 & 11.71 & 12.50 & 11.07 & & &  \\ 
2 & J210144.82+100840.7 & 1RXS J210145.1+100843 & 13.0611 & 9.85 & 10.77 & 9.18 & SAO 126476 & & K2  \\ 
  & J210707.11+063232.1 & 1RXS J210707.1+063247 & 7.1371 & 10.04 & 10.70 & 9.54 & & &  \\ 
  & J211044.78+162323.7 & 1RXS J211044.8+162312 & 10.5750 & 12.34 & 13.62 & 11.26 & & &  \\ 
  & J211436.75+195255.7 & 1RXS J211437.1+195257 & 1.1520 & 11.87 & 13.50 & 11.13 & & &  \\ 
  & J212135.86+094835.3 & 1RXS J212136.2+094834 & 3.6237 & 10.55 & 11.24 & 9.76 & & &  \\ 
  & J212341.65+152148.1 & 1RXS J212342.2+152151 & 6.6225 & 10.37 & 11.42 & 10.10 & & &  \\ 
2 & J212519.57+265653.8 & 1RXS J212519.7+265657 & 2.8246 & 11.26 & 12.20 & 11.11 & & &  \\ 
  & J212812.94+075227.2 & 1RXS J212813.2+075229 & 29.7822 & 12.72 & 13.90 & 11.49 & & &  \\ 
  & J212846.86+232012.5 & 1RXS J212846.4+232009 & 4.5407 & 11.15 & 12.69 & 10.38 & StKM 1-1900 & & K4  \\ 
2 & J212934.76+093530.3 & 1RXS J212935.1+093522 & 2.7362 & 11.71 & 12.71 & 10.51 & & &  \\ 
  & J213004.12+120428.8 & 2RXP J213005.4+120426 & 0.5837 & 15.23 & 15.49 & 14.61 & & &  \\ 
  & J213116.74+225357.1 & 2RXP J213116.9+225402 & 1.1268 & 10.05 & 10.67 & 9.71 & & &  \\ 
  & J213221.70+243342.4 & 1RXS J213220.8+243337 & 4.7358 & 11.99 & 13.98 & 11.22 & GJ 4201 & flare star & M3.5  \\ 
  & J214537.36+271110.8 & 1RXS J214539.0+271124 & 1.3729 & 11.30 & 12.25 & 11.00 & & &  \\ 
  & J214742.17+304210.5 & 1RXS J214741.8+304204 & 34.7828 & 10.25 & 11.62 & 9.37 & BD+30 4528 & & K2  \\ 
2 & J214809.40+191012.9 & 1RXS J214810.6+191013 & 1.1598 & 10.64 & 11.83 & 10.04 & & &  \\ 
  & J214916.06+312502.5 & 1RXS J214916.6+312500 & 6.4128 & 10.82 & 11.54 & 10.49 & & &  \\ 
  & J215323.65+173020.3 & 2RXP J215323.6+173019 & 5.2761 & 13.00 & 13.50 & 12.04 & & &  \\ 
  & J220041.59+271513.5 & 1RXS J220042.0+271520 & 0.5235 & 11.37 & 13.06 & 10.57 & & & K  \\ 
  & J220213.95+152014.2 & 1RXS J220212.8+152012 & 1.4751 & 9.74 & 10.72 & 9.37 & & &  \\ 
2 & J220406.39+343305.3 & 1RXS J220407.2+343309 & 0.3717 & 9.90 & 10.26 & 9.63 & BD+33 4417 & double star & F5  \\ 
  & J221844.11+142130.2 & 2RXP J221844.4+142132 & 4.7004 & 13.13 & 13.59 & 11.89 & & &  \\ 
2 & J222228.81+292212.4 & 1RXS J222228.4+292216 & 0.2779 & 10.44 & 11.43 & 10.12 & & &  \\ 
4 & J222229.09+281439.1 & 1RXS J222229.1+281432 & 2.2761 & 9.34 & 10.66 & 9.46 & SAO 90449/SAO 90450 &double star & G0/F8  \\ 
  & J222558.19+210842.0 & 2RXP J222558.7+210837 & 5.0987 & 13.23 & 14.16 & 12.61 & & &  \\ 
  & J222614.44+213209.6 & 2RXP J222613.9+213219 & 2.8894 & 9.94 & 9.36 & 8.33 & BD+20 5152 & & K0  \\ 
  & J222629.93+212314.5 & 2RXP J222629.8+212318 & 1.0101 & 12.77 & 14.25 & 12.20 & & &  \\ 
  & J222803.99+183606.5 & 1RXS J222804.5+183607 & 0.3232 & 9.97 & 10.35 & 9.72 & BD+17 4751 & & F8  \\ 
  & J222820.71+173959.3 & 1RXS J222819.9+174025 & 2.3142 & 12.02 & 13.38 & 11.92 & & &  \\ 
  & J222829.08+203637.0 & 1RXS J222834.4+203647 & 0.1247 & 13.27 & 13.99 & 12.49 & & &  \\ 
  & J223616.76+331856.7 & 1RXS J223616.0+331909 & 0.3230 & 10.66 & 11.20 & 10.21 & & &  \\ 
  & J223655.17+401027.8 & 2RXP J223655.4+401024 & 15.2555 & 11.42 & 12.18 & 10.87 & & &  \\
  & J224355.18+293647.6 & 1RXS J224353.3+293633 & 0.4443 & 12.54 & 14.15 & 11.42 & & &  \\
  & J224446.13+302933.6 & 1RXS J224446.2+302927 & 3.5533 & 10.40 & 11.07 & 9.80 & & &  \\
2 & J225155.50+353915.2 & 1RXS J225155.6+353911 & 5.2276 & 11.03 & 12.24 & 10.69 & & &  \\
  & J225338.13+291305.0 & 2RXP J225337.7+291310 & 0.8252 & 12.33 & 12.74 & 11.28 & & &  \\
  & J225454.99+241445.2 & 1RXS J225453.7+241449 & 0.1994 & 12.70 & 15.68 & 12.91 & & &  \\
2 & J225538.93+281035.2 & 1RXS J225537.7+281051 & 0.8442 & 11.77 & 12.60 & 11.44 & & &  \\
  & J225617.59+205236.2 & 2RXP J225617.9+205236 & 1.0999 & 11.60 & 12.84 & 10.95 & & &  \\
2 & J225849.98+405611.4 & 1RXS J225850.4+405610 & 6.3266 & 10.86 & 11.66 & 10.34 & & &  \\
  & J225923.53+325133.3 & 1RXS J225923.2+325127 & 2.3833 & 11.90 & 12.98 & 11.24 & & &  \\
  & J230147.77+352848.3 & 1RXS J230147.6+352854 & 14.1358 & 9.74 & 10.86 & 9.03 & SAO 72938 & & K2  \\
2 & J230209.26+351539.4 & 1RXS J230209.2+351538 & 9.1534 & 10.05 & 11.01 & 9.51 & & &  \\
  & J230724.88+315014.1 & 1RXS J230725.0+315012 & 7.7129 & 10.67 & 12.08 & 9.73 & & &  \\
  & J230843.02+213717.7 & 1RXS J230842.4+213716 & 7.5465 & 9.40 & 10.11 & 8.80 & SAO 91038 & & G5  \\
  & J231036.91+205526.2 & 1RXS J231037.8+205531 & 0.9198 & 10.16 & 10.87 & 9.37 & HD 218782 & & K2  \\
  & J231059.54+214243.2 & 1RXS J231100.1+214254 & 0.2582 & 11.22 & 12.51 & 11.01 & & &  \\
2 & J231206.52+265545.7 & 1RXS J231206.8+265552 & 0.1563 & 9.37 & 13.10 & 7.39 & HDS 3305 & double star & K3  \\
  & J231455.84+273958.5 & 1RXS J231456.5+273957 & 11.7668 & 11.59 & 12.62 & 10.52 & & &  \\
  & J232048.10+292155.6 & 1RXS J232048.8+292151 & 1.3710 & 11.52 & 11.97 & 10.86 & & &  \\
2 & J232153.08+231656.2 & 1RXS J232153.8+231703 & 18.8148 & 10.90 & 11.59 & 10.43 & & &  \\
  & J232617.06+275203.5 & 1RXS J232617.4+275200 & 1.2530 & 11.72 & 13.67 & 11.11 & & &  \\
4 & J233152.17+195614.2 & 1RXS J233152.6+195735 & 1.0664 & 9.63 & 12.12 & 9.55 & EQ Peg & UV Cet  & M3.5  \\
  & J233906.78+220412.4 & 1RXS J233907.3+220355 & 4.0623 & 10.55 & 11.42 & 9.79 & BD+21 4966 & & K2  \\
  & J234028.98+295911.8 & 1RXS J234027.8+295912 & 1.1648 & 11.16 & 10.48 & 8.90 & & &  \\
  & J234106.15+270643.2 & 2RXP J234106.2+270638 & 9.8987 & 11.95 & 12.79 & 10.77 & & &  \\
  & J234720.80+300510.8 & 1RXS J234721.7+300505 & 2.3252 & 10.81 & 11.49 & 10.41 & & &  \\
  & J234945.36+312627.4 & 1RXS J234944.9+312629 & 1.3808 & 12.77 & 13.56 & 11.60 & & &  \\
  & J235750.15+334348.5 & 1RXS J235750.3+334401 & 1.4194 & 11.21 & 12.66 & 10.96 & & &  \\
  & J235952.73+294947.3 & 2RXP J235952.7+294949 & 0.5657 & 13.22 & 13.70 & 12.15 & & &  \\ \hline
\end{longtable}
\normalsize

\end{document}